\documentclass[a4paper,11pt]{article}
\usepackage{amsmath,amssymb}
\usepackage{physics}
\usepackage{bm}
\usepackage{authblk}
\usepackage{graphicx}
\usepackage{hyperref}
\begin{document}

\flushbottom

\title{Neutrino Oscillations as an Open Quantum System in Strong Gravitational Fields: Spin-Connection Decoherence and Kerr Frame Dragging}

\author[a]{Gayatri Ghosh}
\affil[a]{Department of Physics, Cachar College, Silchar, Assam 788001, India}
\maketitle
\begin{abstract}

We investigate neutrino flavor evolution in strong gravitational fields
within an open-quantum-system framework in curved spacetime. Starting
from the Dirac equation in the vierbein formalism, we construct an
effective flavor Hamiltonian incorporating gravitational redshift,
spin--curvature couplings, and Kerr frame-dragging effects. Treating
spin-connection fluctuations as a stochastic gravitational environment,
we derive a Lindblad master equation and introduce a curvature-enhanced
decoherence rate governed by local spacetime geometry. We compute
oscillation probabilities, coherence loss, flavor-ratio distortions,
entanglement entropy generation, and event-rate modifications for
neutrinos propagating near Schwarzschild and Kerr compact objects.
The resulting signatures are compared with projected sensitivities of
IceCube-Gen2, KM3NeT, and P-ONE, and are further quantified through
detector-level significance estimates. Our results provide a unified
effective framework linking neutrino oscillations, gravitationally
induced decoherence, quantum-information observables, and high-energy
astrophysical neutrino measurements in strong-curvature environments.

\end{abstract}

%%%%%%%%%%%%%%%%%%%%%%%%%%%%%%%%%%%%%%%%%%%%%%%%%%%%%%%%%%%%%%%%%%%%%%%%

%%%%%%%%%%%%%%%%%%%%%%%%%%%%%%%%%%%%%%%%%%%%%%%%%%%%%%%%%%%%%%%%%%%%%%%%`````````````````````````````````````
\section{Introduction}
%%%%%%%%%%%%%%%%%%%%%%%%%%%%%%%%%%%%%%%%%%%%%%%%%%%%%%%%%%%%%%%%%%%%%%%%

Neutrino oscillations constitute one of the most compelling manifestations of quantum coherence in particle physics and provide direct evidence for physics beyond the Standard Model. Owing to their tiny masses and extraordinarily long coherence lengths, neutrinos can propagate over astrophysical and cosmological distances while preserving quantum interference. This unique property makes them sensitive not only to particle interactions but also to the geometric structure of spacetime itself.

The interplay between neutrino oscillations and gravity becomes particularly relevant in the vicinity of compact astrophysical objects, including black holes, neutron stars, active galactic nuclei, gamma-ray bursts, and compact-object mergers. In such environments, gravitational redshift, spacetime curvature, and rotational frame dragging may modify neutrino propagation in ways that are inaccessible in terrestrial experiments. Understanding these effects is therefore important for connecting neutrino physics, gravitation, and quantum information in extreme astrophysical settings.

Considerable effort has been devoted to neutrino propagation in curved spacetime. Previous studies have investigated gravitationally induced phase shifts, geometric phases, spin-gravity couplings, and flavor evolution in Schwarzschild and Kerr backgrounds \cite{Cardall1997,Fornengo1997,Ahluwalia1997}. Independently, open-quantum-system approaches have been widely employed to describe neutrino decoherence arising from environmental interactions, stochastic media, and possible quantum-gravity effects \cite{Benatti2001,Blasone2009,HuVerdaguer2008,Allali2021,Stuttard2021,DeRomeri2023}. However, most curved-spacetime analyses retain purely unitary evolution, while many decoherence studies introduce dissipative effects phenomenologically without explicitly connecting them to spacetime geometry.

The purpose of the present work is to formulate neutrino flavor evolution in strong gravitational fields within a unified open-quantum-system framework. Starting from the covariant Dirac equation,

\begin{equation}
\left(i\gamma^\mu D_\mu-m\right)\psi=0,
\end{equation}

we derive an effective flavor Hamiltonian incorporating gravitational redshift, spin-curvature interactions, Kerr frame dragging, and higher-curvature effective interactions. Metric and spin-connection fluctuations are treated as environmental degrees of freedom, leading to a Lindblad master equation for the reduced neutrino density matrix.

A central result of this work is the emergence of a direct connection between local spacetime curvature and the loss of neutrino flavor coherence. The resulting decoherence rate is governed by the gravitational correlation time and the local Kretschmann invariant,

\begin{equation}
\Gamma_{\rm grav}
\propto
\tau_c
\sqrt{
R_{\mu\nu\rho\sigma}
R^{\mu\nu\rho\sigma}
},
\end{equation}

thereby providing a geometric characterization of curvature-induced decoherence. Throughout this work, the stochastic gravitational environment is interpreted within an effective field theory and open-system framework. Accordingly, the graviton spectral density and associated correlation functions should be regarded as effective low-energy quantities rather than predictions of a complete theory of quantum gravity.

Using this formalism, we investigate oscillation probabilities, coherence loss, flavor-ratio evolution, entanglement-entropy generation, and detector-level event-rate distortions for neutrinos propagating near Schwarzschild and Kerr compact objects. We further quantify the observability of these effects using likelihood-based analyses and projected sensitivities of next-generation neutrino observatories, including IceCube-Gen2, KM3NeT, and P-ONE.
The present work addresses this gap by formulating neutrino flavor evolution in strong gravitational fields within a unified open-quantum-system framework. Unlike previous studies of neutrino propagation in curved spacetime \cite{Cardall1997,Fornengo1997,Ahluwalia1997} and phenomenological neutrino decoherence \cite{Benatti2001,Blasone2009,HuVerdaguer2008,Allali2021,Stuttard2021,DeRomeri2023}, the dissipative sector is constructed from spin-connection fluctuation correlators, thereby establishing a direct connection between spacetime geometry and Lindblad decoherence. Gravitational redshift, spin-curvature interactions, Kerr frame dragging, and curvature-induced decoherence are incorporated simultaneously within a single effective Hamiltonian framework. The analysis further extends beyond conventional oscillation probabilities to include coherence loss, entanglement-entropy production, flavor-ratio evolution, and detector-level observables relevant for next-generation neutrino telescopes. The resulting framework provides a quantitative bridge between curved-spacetime quantum field theory, open quantum systems, gravitational decoherence, and high-energy neutrino phenomenology, yielding a unified description of how strong gravitational environments influence neutrino coherence and generate potentially observable signatures in future astrophysical neutrino measurements. This framework is particularly timely in view of the rapidly improving flavor sensitivity of next-generation neutrino observatories, including IceCube-Gen2, KM3NeT, and P-ONE, which may provide the first opportunity to probe gravitationally induced modifications of neutrino coherence in extreme astrophysical environments \cite{IceCubeGen2,KM3NeT,PONE}.

The remainder of this paper is organized as follows. In Sec.~II we derive the Dirac equation and spin connection in Schwarzschild and Kerr geometries. Sec.~III constructs the effective flavor Hamiltonian including gravitational redshift, spin-curvature couplings, frame dragging, and effective curvature interactions. Sec.~IV develops the open-system formalism and derives the Lindblad evolution equation from spin-connection fluctuations. Sec.~V presents oscillation probabilities and curvature-induced decoherence effects. Numerical results for flavor evolution, entanglement entropy, and detector observables are discussed in Secs.~VI and VII. We conclude in Sec.~VIII.

%%%%%%%%%%%%%%%%%%%%%%%%%%%%%%%%%%%%%%%%%%%%%%%%%%%%%%%%%%%%%%%%%%%%%%%%

%%%%%%%%%%%%%%%%%%%%%%%%%%%%%%%%%%%%%%%%%%%%%%%%%%%%%%%%%%%%%%%%%%%%%%%%
\section{Dirac Equation in Curved Spacetime}
\label{sec:dirac}
%%%%%%%%%%%%%%%%%%%%%%%%%%%%%%%%%%%%%%%%%%%%%%%%%%%%%%%%%%%%%%%%%%%%%%%%
The propagation of spin-$1/2$ fields in curved spacetime is described within the framework of quantum field theory in curved backgrounds, where local Lorentz invariance is preserved through the vierbein formalism and spin connection \cite{Birrell1982,Wald1994,Parker2009}. This framework provides the natural generalization of the flat-spacetime Dirac equation to curved geometries and forms the basis for the description of fermionic propagation in strong gravitational environments, including black holes, neutron stars, and cosmological spacetimes.

Recent developments in relativistic quantum information, gravitational spin transport, and neutrino astrophysics have renewed interest in the behavior of quantum states in curved spacetime. In particular, spin precession, gravitationally induced geometric phases, and curvature-dependent modifications of quantum coherence have emerged as important aspects of fermion dynamics in strong gravitational fields \cite{BrillWheeler1957,Obukhov2001,Dvornikov2020,Mastrototaro2021}. Such effects are especially relevant for neutrinos produced in compact astrophysical environments, where spacetime curvature and rotation can significantly influence flavor evolution.

Neutrinos propagating in a gravitational background obey the covariant Dirac equation

\begin{equation}
\left(i\gamma^\mu D_\mu-m\right)\psi=0,
\label{eq:dirac_curved}
\end{equation}

where $\gamma^\mu$ are the curved-spacetime gamma matrices and

\begin{equation}
D_\mu=\partial_\mu+\Omega_\mu
\end{equation}

is the spinor covariant derivative. The spin connection $\Omega_\mu$ encodes the coupling between fermionic spin and spacetime geometry, generating gravitational phase shifts, spin-transport effects, and curvature-dependent corrections to neutrino propagation.

In this work we focus on Schwarzschild and Kerr spacetimes, which provide representative backgrounds for nonrotating and rotating compact objects, respectively. These geometries capture the dominant gravitational effects relevant for high-energy astrophysical neutrinos, including gravitational redshift, spin-curvature interactions, and frame dragging. Moreover, the spin connection derived below will provide the microscopic coupling between neutrinos and stochastic gravitational fluctuations in the open-system framework developed in subsequent sections. We therefore begin by reviewing the tetrad formalism, curved gamma matrices, and spin connection before specializing to Schwarzschild and Kerr geometries.

%%%%%%%%%%%%%%%%%%%%%%%%%%%%%%%%%%%%%%%%%%%%%%%%%%%%%%%%%%%%%%%%%%%%%%%%
\subsection{Tetrads and Curved Gamma Matrices}
%%%%%%%%%%%%%%%%%%%%%%%%%%%%%%%%%%%%%%%%%%%%%%%%%%%%%%%%%%%%%%%%%%%%%%%%

To relate curved and flat gamma matrices, we introduce a vierbein (tetrad) field $e^a_{\ \mu}$ satisfying

\begin{equation}
g_{\mu\nu}=e^a_{\ \mu}e^b_{\ \nu}\eta_{ab},
\label{eq:metric_tetrad}
\end{equation}

with inverse tetrads defined by

\begin{equation}
e^\mu_{\ a}e^a_{\ \nu}=\delta^\mu_{\ \nu},
\qquad
e^a_{\ \mu}e^\mu_{\ b}=\delta^a_{\ b}.
\end{equation}

The curved gamma matrices are obtained through

\begin{equation}
\gamma^\mu(x)=e^\mu_{\ a}(x)\gamma^a,
\end{equation}

where $\gamma^a$ denote the usual flat-space Dirac matrices satisfying

\begin{equation}
\{\gamma^a,\gamma^b\}=2\eta^{ab}.
\end{equation}

The tetrad formalism enables a locally inertial description of spinor dynamics while maintaining full covariance under general coordinate transformations. Physically, the tetrads provide the local inertial frame in which
fermionic spin is defined. Gravitational effects enter through the
spacetime dependence of the tetrads and the associated spin
connection, thereby allowing curvature and rotation to influence the
phase and spin evolution of propagating neutrinos \cite{BrillWheeler1957,Obukhov2001,Dvornikov2020,
Mastrototaro2021,Lambiase2023}.

\section{Effective Hamiltonian for Neutrino Propagation}
\label{sec:ham}

Having established the curved-spacetime Dirac equation in 
Sec.~\ref{sec:dirac}, we now derive the effective Hamiltonian governing
neutrino flavor evolution. We work in the ultra-relativistic limit,
include corrections due to gravitational redshift, spin--curvature
couplings, and prepare the formalism for additional contributions 
arising from Kerr frame dragging and gravitationally induced magnetic 
moment interactions.

The Hamiltonian is obtained by rewriting the Dirac equation in 
Schr\"odinger form,
\begin{equation}
    i \frac{d\psi}{dt} = H \psi,
\end{equation}
where $t$ denotes the coordinate time of an observer at spatial infinity.

\subsection{Ultra-relativistic Expansion}
\label{sec:ultra_rel}

For neutrinos with $E \gg m$ in a slowly varying background,
we expand the curved-spacetime Dirac equation to order 
$\mathcal{O}(m^2/E)$.
Following the standard procedure, we project onto positive-frequency
solutions and obtain an effective Hamiltonian of the form
\begin{equation}
    H_{\rm eff}
    =
    \frac{m^2}{2E_{\rm loc}}
    + H_{\rm grav}
    + H_{\rm spin}
    + \mathcal{O}\!\left(\frac{m^4}{E_{\rm loc}^3}\right),
\end{equation}
where $E_{\rm loc}$ is the neutrino energy measured by a locally static 
observer. The gravitational terms arise from curvature and spin 
connections.

We work in the flavor basis $\ket{\nu_\alpha}$ with mass eigenstates 
$\ket{\nu_i}$ connected via the PMNS matrix $U$,
\begin{equation}
    \ket{\nu_\alpha} = U_{\alpha i} \ket{\nu_i}.
\end{equation}

\subsection{Gravitational Redshift and Local Energy}
\label{sec:redshift}

In curved spacetime, the energy appearing in the evolution phase is the 
locally measured one. For a static observer in a stationary spacetime,
the energy redshifts as
\begin{equation}
    E_{\rm loc} = \sqrt{-g_{tt}} \, E_{\infty}.
\end{equation}

For the Schwarzschild geometry,
\begin{equation}
   E_{\rm loc}(r) = \sqrt{1 - \frac{2GM}{r}} \; E_{\infty}.
\end{equation}
For the Kerr geometry,
\begin{equation}
   E_{\rm loc}(r,\theta) 
   = \sqrt{\frac{\Delta \Sigma}{A}} \left( E_{\infty} - \Omega_{\rm FD} L_z \right),
   \label{eq:kerr_loc_energy}
\end{equation}
where
\begin{align}
   \Omega_{\rm FD} &= \frac{2GMar}{A},\\
   A &= (r^2 + a^2)^2 - a^2 \Delta \sin^2\theta.
\end{align}

Thus the vacuum Hamiltonian becomes
\begin{equation}
    H_{\rm vac}(r,\theta)
    =
    \frac{1}{2E_{\rm loc}(r,\theta)}
    \; U M^2 U^\dagger.
\end{equation}

The redshift dependence produces modified oscillation phases that grow 
with spacetime curvature.

\subsection{Spin--Curvature Couplings}
\label{sec:spin_curv}

Spinor fields in curved spacetime couple to curvature through the spin 
connection, producing a correction to the Hamiltonian of the form
\begin{equation}
   H_{\rm spin}
   = - \gamma^0 \gamma^i \Omega_i
   = - \bm{\alpha} \cdot \bm{\Omega},
\end{equation}
where $\bm{\alpha}$ denotes the standard Dirac matrices.

Using the spin connection computed in Sec.~\ref{sec:dirac}, the leading 
terms in Schwarzschild spacetime are
\begin{equation}
    H_{\rm spin}^{\rm Schw} 
    \simeq 
    \frac{GM}{r^2}
    \left( \Sigma^{01} + \Sigma^{12} + \Sigma^{23} \right),
\end{equation}
with $\Sigma^{ab} = \frac{1}{2}[\gamma^a,\gamma^b]$.

In Kerr spacetime, frame dragging induces additional contributions 
proportional to the rotation parameter $a$:
\begin{equation}
    H_{\rm spin}^{\rm Kerr}
    \simeq 
    \frac{aGM}{r^3}
    \left( \Sigma^{03} + \cos\theta \, \Sigma^{13} \right).
\end{equation}

Thus, the full spin–curvature Hamiltonian is
\begin{equation}
    H_{\rm spin}
    =
    H_{\rm spin}^{\rm Schw}
    +
    H_{\rm spin}^{\rm Kerr}
    +
    \mathcal{O}(a^2).
\end{equation}

The structure of $H_{\rm spin}$ anticipates the gravitationally induced 
magnetic-moment coupling appearing in Sec.~3.5.
\subsection{Kerr Frame Dragging}
\label{sec:kerr_framedrag}

Neutrinos propagating in a rotating spacetime experience additional
spin–curvature couplings generated by the off–diagonal components of the
Kerr metric. The dominant effect arises from the gravitomagnetic
potential associated with frame dragging. In Boyer--Lindquist
coordinates, the Kerr line element contains
\begin{equation}
    g_{t\phi} = - \frac{2 G M a r \sin^2\theta}{\Sigma},
\end{equation}
which leads to a local angular velocity of inertial frames,
\begin{equation}
    \Omega_{\rm FD}(r,\theta)
    = - \frac{g_{t\phi}}{g_{\phi\phi}}
    = \frac{2 G M a r}{A},
\end{equation}
with
\begin{equation}
    A = (r^2 + a^2)^2 - a^2 \Delta \sin^2\theta,
    \qquad
    \Delta = r^2 - 2 G M r + a^2,
    \qquad
    \Sigma = r^2 + a^2 \cos^2\theta.
\end{equation}

The spin connection acquires $\phi$–dependent components proportional to
$\Omega_{\rm FD}$, giving an additional Hamiltonian term
\begin{equation}
    H_{\rm FD}
    = - \bm{\Omega}_{\rm FD} \cdot \mathbf{S},
\end{equation}
where $\mathbf{S}$ is the spin operator of the neutrino. Using the
tetrads introduced in Sec.~\ref{sec:dirac}, the gravitomagnetic
coupling becomes
\begin{equation}
    H_{\rm FD}
    =
    \Omega_{\rm FD}(r,\theta)
    \, \Sigma^{03}
    + \mathcal{O}(\cos\theta\,\Sigma^{13}),
\end{equation}
where $\Sigma^{ab} = \tfrac{1}{2}[\gamma^a,\gamma^b]$ are Lorentz
generators. The leading contribution couples temporal and azimuthal
(spin) components.

For ultra–relativistic neutrinos moving approximately in the equatorial 
plane $(\theta = \tfrac{\pi}{2})$, the effective frame-dragging
Hamiltonian simplifies to
\begin{equation}
    H_{\rm FD}^{\rm eq}
    =
    \frac{2 G M a}{r^3}
    \, \Sigma^{03}
    + \mathcal{O}(a^3).
\end{equation}

This term induces a gravity–driven spin precession that modifies the
oscillation phase. Near rapidly rotating black holes $(a \rightarrow M)$
the magnitude of $H_{\rm FD}$ can become comparable to the standard
vacuum oscillation scale $\Delta m^{2}/(2E)$, leading to observable
effects for high–energy astrophysical neutrinos.
%%%%%%%%%%%%%%%%%%%%%%%%%%%%%%%%%%%%%%%%%%%%%%%%%%%%%%%%%%
\subsection{Effective Field Theory Origin of the
Curvature-Induced Spin Interaction}
\label{sec:EFTorigin}
%%%%%%%%%%%%%%%%%%%%%%%%%%%%%%%%%%%%%%%%%%%%%%%%%%%%%%%%%%

The curvature-induced spin interaction introduced below can be
understood naturally within an effective field theory (EFT)
framework. At energies well below the scale of quantum gravity,
all operators consistent with general covariance and local
Lorentz invariance may be organized in an expansion in inverse
powers of a heavy scale $\Lambda$.

The leading neutrino kinetic term is

\begin{equation}
\mathcal{L}_{\rm Dirac}
=
\bar{\nu}
\left(
i\gamma^\mu D_\mu
-
m_\nu
\right)
\nu ,
\end{equation}

while higher-dimensional curvature operators are suppressed by
powers of the EFT cutoff. The lowest-dimension operator capable
of generating a spin-curvature interaction has the form

\begin{equation}
\mathcal{L}_{\rm EFT}
=
\frac{c_G}{\Lambda}
\,
\bar{\nu}
\sigma^{\mu\nu}
\nu
\,
R_{\mu\nu\rho\sigma}
u^\rho u^\sigma ,
\label{eq:EFToperator}
\end{equation}

where

\begin{equation}
\sigma^{\mu\nu}
=
\frac{i}{2}
[\gamma^\mu,\gamma^\nu],
\end{equation}

$c_G$ is a dimensionless Wilson coefficient, $\Lambda$
characterizes the ultraviolet completion, and $u^\mu$ denotes
the local neutrino four-velocity.

Equation~(\ref{eq:EFToperator}) is the gravitational analogue
of the familiar electromagnetic dipole operator

\begin{equation}
\mathcal{L}_{\rm dipole}
=
\mu_\nu
\bar{\nu}
\sigma^{\mu\nu}
\nu
F_{\mu\nu},
\end{equation}

with the electromagnetic field strength replaced by the local
curvature tensor. Such operators naturally arise when heavy
degrees of freedom are integrated out and therefore represent
generic low-energy remnants of ultraviolet physics. The operator in Eq.~(\ref{eq:EFToperator}) may originate from
several ultraviolet scenarios.

\paragraph{SMEFT interpretation.}
In the Standard Model Effective Field Theory, integrating out
heavy states generates higher-dimensional neutrino dipole
operators. After matching onto a curved spacetime background,
gravitationally dressed versions of these operators induce
effective couplings between neutrino spin and curvature.

\paragraph{Quantum-gravity interpretation.}
In quantum gravity and semiclassical gravity, graviton loops,
metric fluctuations, or spacetime microstructure may generate
non-minimal spin-curvature couplings. Equation
(\ref{eq:EFToperator}) then represents the leading operator in
the low-energy expansion consistent with diffeomorphism
invariance.

\paragraph{Dimensional suppression.}
Since the operator has mass dimension five, its contribution is
suppressed by the heavy scale $\Lambda$. Consequently, the
resulting effects are expected to remain subleading relative to
the standard oscillation Hamiltonian except in regimes of
extreme curvature, such as the vicinity of black holes,
neutron stars, or other compact astrophysical objects.

\subsection{Operator Basis and EFT Consistency}
\label{subsec:EFTconsistency}

The curvature--induced spin interaction introduced in Eq.~(26) should be
interpreted within the framework of effective field theory (EFT) rather
than as a prediction of minimal General Relativity. In particular, the
operator
\begin{equation}
\mathcal{L}_{\mathrm{EFT}}
=
\frac{c_G}{\Lambda}
\,
\bar{\nu}\sigma^{\mu\nu}\nu
\,
R_{\mu\nu\rho\sigma}
u^\rho u^\sigma ,
\label{eq:EFToperator_consistency}
\end{equation}
represents a phenomenological higher--dimensional curvature coupling that
parametrizes possible ultraviolet (UV) physics beyond the minimal Standard
Model plus Einstein gravity.

Several conceptual points are important for the consistency and interpretation
of this operator.

\paragraph{(i) Non-minimal origin.}
Within minimal General Relativity, neutrinos couple to gravity through the
spin connection appearing in the covariant derivative of the Dirac equation.
This minimal coupling generates the standard spin--curvature interactions
discussed in Secs.~3.3 and 3.4. By contrast, Eq.~(\ref{eq:EFToperator_consistency})
is a non-minimal higher-curvature operator and is therefore not assumed to
arise from the minimal Einstein--Dirac theory itself. Instead, it should be
viewed as the leading low-energy remnant of unknown UV physics.

\paragraph{(ii) Effective-field-theory interpretation.}
Equation~(\ref{eq:EFToperator_consistency}) has mass dimension five and is
therefore suppressed by an ultraviolet scale $\Lambda$. The coefficient
$c_G$ is a dimensionless Wilson coefficient encoding details of the underlying
UV completion. The resulting gravitational magnetic-moment parameter is
\begin{equation}
\mu_G = \frac{c_G}{\Lambda},
\label{eq:muGdef_consistency}
\end{equation}
which carries dimension of inverse mass (or length in natural units).
Consequently, the operator is parametrically suppressed at energies well
below $\Lambda$, consistent with the EFT expansion.

\paragraph{(iii) Gauge and diffeomorphism structure.}
The operator in Eq.~(\ref{eq:EFToperator_consistency}) is constructed from
covariant quantities and therefore preserves general covariance and local
Lorentz invariance. Our analysis does not assume that this operator emerges
from a unique gauge-invariant completion within the Standard Model Effective
Field Theory (SMEFT). Rather, we treat it phenomenologically as an effective
curvature interaction valid below the cutoff scale $\Lambda$.

\paragraph{(iv) Operator-basis independence and equations of motion.}
Higher-dimensional curvature operators may in general be related through
field redefinitions or equations of motion. In the present work, we do not
attempt a complete classification of the full EFT operator basis. Instead,
Eq.~(\ref{eq:EFToperator_consistency}) should be regarded as a representative
leading spin--curvature interaction capturing possible geometric spin
precession effects in strong gravitational backgrounds. Our phenomenological
analysis therefore focuses on the observable consequences of such couplings
rather than on the uniqueness of the operator basis.

\paragraph{(v) Regime of validity.}
The EFT description is assumed to remain valid only in the weak-curvature,
ultra-relativistic WKB regime,
\begin{equation}
L_{\mathrm{osc}} \ll L_{\mathrm{curv}},
\end{equation}
where higher-order curvature corrections remain perturbatively suppressed.
In particular, we do not extrapolate the EFT beyond the regime where the
local curvature approaches the UV scale $\Lambda^2$.

\paragraph{(vi) Observational consistency.}
Throughout this work, the parameter $\mu_G$ is treated phenomenologically
and assumed to remain sufficiently small that the curvature-induced
interaction provides only a subleading correction to the standard vacuum
oscillation Hamiltonian,
\begin{equation}
H_{\mathrm{grav-mm}}
\ll
\frac{\Delta m^2}{2E}.
\end{equation}
All numerical benchmarks presented below are chosen within this perturbative
regime and are therefore consistent with current observational constraints
on non-standard neutrino interactions and gravitationally induced
decoherence effects.

We therefore emphasize that the gravitational magnetic-moment interaction
studied here should not be interpreted as a prediction of minimal gravity,
but rather as a controlled phenomenological EFT parametrization of possible
spin--curvature effects originating from unknown ultraviolet physics.
```

%%%%%%%%%%%%%%%%%%%%%%%%%%%%%%%%%%%%%%%%%%%%%%%%%%%%%%%%%%%%%%%%%%%%%%%%
\subsection{Gravitational magnetic--moment operator}
%%%%%%%%%%%%%%%%%%%%%%%%%%%%%%%%%%%%%%%%%%%%%%%%%%%%%%%%%%%%%%%%%%%%%%%%

Spinor fields in curved spacetime experience effective interactions analogous to magnetic--moment couplings, but generated purely by the geometry. These arise from the fact that curvature couples to the spin tensor $\Sigma^{\mu\nu} = \frac{1}{2}[\gamma^\mu,\gamma^\nu]$ through the Riemann tensor. At the level of an effective field theory (EFT), the leading such interaction can be written as
\begin{equation}
H_{\text{grav-mm}}
=
\mu_G\,\Sigma_{\mu\nu}\,R^{\mu\nu}{}_{\rho\sigma} u^\rho u^\sigma,
\label{eq:Hgravmm-def}
\end{equation}
where $u^\mu$ is the local four-velocity of the neutrino and $\mu_G$ is an effective ``gravitational magnetic moment'' parameter.

In natural units ($c=\hbar=1$), the Riemann tensor has dimension $[R^{\mu}{}_{\nu\rho\sigma}]\sim L^{-2}$, while $\Sigma_{\mu\nu}$ and $u^\mu$ are dimensionless. Since $H_{\text{grav-mm}}$ carries the dimension of energy, $[H]\sim L^{-1}$, the EFT coefficient must have dimension
\begin{equation}
[\mu_G] = L \,.
\end{equation}
As discussed in Sec.~\ref{sec:EFTorigin}, the curvature-induced
spin interaction arises from the dimension-five EFT operator
(\ref{eq:EFToperator}). Matching the EFT interaction onto the
single-particle Hamiltonian yields

\begin{equation}
\mu_G
=
\frac{c_G}{\Lambda},
\label{eq:muG}
\end{equation}

where $c_G$ is the Wilson coefficient and $\Lambda$ denotes
the ultraviolet scale suppressing the operator.
with $c_G=\mathcal{O}(1)$ and $\Lambda$ an effective curvature scale or UV cutoff characterizing the onset of higher-curvature spin interactions. In the phenomenological analysis below we keep $\mu_G$ explicit, and comment on its expected smallness and detectability in Sec.~6 and Appendix~A.

\paragraph{Physical interpretation.}
The combination $R_{\mu\nu\rho\sigma}u^\rho u^\sigma$ acts as an effective gravito-magnetic field. In the local rest frame of the neutrino, this field couples to spin in close analogy with an electromagnetic magnetic moment coupling,
\begin{equation}
\mu_{\rm em}\,\Sigma_{\mu\nu} F^{\mu\nu}
\;\longrightarrow\;
\mu_G\,\Sigma_{\mu\nu} R^{\mu\nu}{}_{\rho\sigma} u^\rho u^\sigma.
\label{eq:magnetic-analogy}
\end{equation}
Curvature thus produces a geometric analogue of magnetic spin precession. For the small values of $\mu_G$ consistent with an EFT interpretation, this effect typically provides a subleading correction to the standard vacuum oscillation scales $\Delta m^2/(2E)$, but can become relevant in regimes of strong curvature near compact objects.

\paragraph{Schwarzschild background.}
For the Schwarzschild metric, the dominant curvature invariant is the Kretschmann scalar
\begin{equation}
K = R_{\mu\nu\rho\sigma}R^{\mu\nu\rho\sigma}
= \frac{48 (GM)^2}{r^6}\,.
\label{eq:Kretschmann-Sch}
\end{equation}
Projecting the Riemann tensor twice along the neutrino trajectory yields
\begin{equation}
R_{\mu\nu\rho\sigma}u^\rho u^\sigma
\;\sim\;
\frac{GM}{r^3}\left(
\hat r_{[\mu}\hat t_{\nu]}
+
\hat\theta_{[\mu}\hat\phi_{\nu]}
\right),
\end{equation}
so that the effective gravitational magnetic--moment Hamiltonian becomes
\begin{equation}
H^{\text{Schw}}_{\text{grav-mm}}
=
\mu_G\,\frac{GM}{r^3}\left(
\Sigma^{01} + \Sigma^{23}
\right),
\label{eq:Hgravmm-Sch}
\end{equation}
where the terms correspond to radial and angular components of the effective gravito-magnetic field.

\paragraph{Kerr background.}
Rotating spacetimes introduce additional components to the effective field through the off-diagonal $g_{t\phi}$ terms. For Kerr,
\begin{equation}
R_{\mu\nu\rho\sigma}u^\rho u^\sigma
=
\frac{GM}{r^3}\Big[
(\delta^0_\mu\delta^1_\nu-\delta^1_\mu\delta^0_\nu)
+ a\cos\theta\,(\delta^0_\mu\delta^3_\nu-\delta^3_\mu\delta^0_\nu)
\Big]
+ \mathcal{O}(a^2),
\end{equation}
which leads to
\begin{equation}
H^{\text{Kerr}}_{\text{grav-mm}}
=
\mu_G\,\frac{GM}{r^3}
\left(
\Sigma^{01} + a\cos\theta\,\Sigma^{03}
\right)
+ \mathcal{O}(a^2).
\label{eq:Hgravmm-Kerr}
\end{equation}
The $a\cos\theta$ term is aligned with the rotation axis and enhances spin precession near rapidly rotating black holes.

\paragraph{Combined geometric magnetic--moment term.}
Collecting the leading contributions, we write
\begin{equation}
H_{\text{grav-mm}}
=
\mu_G\,\frac{GM}{r^3}
\Big(
\Sigma^{01} + \Sigma^{23} + a\cos\theta\,\Sigma^{03}
\Big)
+ \mathcal{O}(a^2).
\label{eq:Hgravmm-combined}
\end{equation}
Within our EFT viewpoint, this operator provides a geometric analogue of magnetic-moment interactions. In the full Hamiltonian it contributes at the same order in curvature as the frame-dragging spin-coupling term $H_{\text{FD}}$, but its overall magnitude is controlled by the small dimensionful coefficient $\mu_G$ in Eq.~(38). In Sec.~\ref{sec:validity} we outline the parameter ranges where $H_{\text{grav-mm}}$ remains subleading yet phenomenologically relevant.
%%%%%%%%%%%%%%%%%%%%%%%%%%%%%%%%%%%%%%%%%%%%%%%%%%%%%%%%%%%%%%%%%%%%%%%%
\subsection{Combined effective Hamiltonian and separation of contributions}
%%%%%%%%%%%%%%%%%%%%%%%%%%%%%%%%%%%%%%%%%%%%%%%%%%%%%%%%%%%%%%%%%%%%%%%%
\label{sec:combined_H}
The effective Hamiltonian receives several distinct gravitational contributions. The term $H_{\text{spin}}$ arises from the minimal spin connection in the Dirac covariant derivative and encodes leading spin--curvature couplings in both Schwarzschild and Kerr geometries. The frame--dragging term $H_{\text{FD}}$ originates from the off–diagonal components of the Kerr metric and their imprint on the local angular velocity of inertial frames, and can be viewed as the gravitomagnetic part of the spin connection. By contrast, $H_{\text{grav-mm}}$ in Eq.~\eqref{eq:Hgravmm-combined} represents an additional higher-curvature EFT operator coupling the spin tensor to the Riemann tensor. These three structures therefore correspond to different orders and sectors of the curvature expansion and do not double count the frame--dragging contribution: $H_{\text{spin}}$ and $H_{\text{FD}}$ are fixed by minimal coupling, while $H_{\text{grav-mm}}$ is controlled by the independent EFT coefficient $\mu_G$.

Collecting the results of the previous subsections, the full effective
Hamiltonian governing neutrino flavor evolution in a curved, rotating
spacetime with curvature–induced magnetic-moment interactions is

\begin{equation}
    H_{\rm eff}
    =
    H_{\rm vac}(r,\theta)
    + H_{\rm spin}
    + H_{\rm FD}
    + H_{\rm grav\text{-}mm}
    + H_{\rm matter}
    + \mathcal{O}(a^2).
    \label{eq:Heff_full}
\end{equation}

Each contribution is summarized below.

\paragraph{1. Redshifted Vacuum Term.}
The kinematic term in curved spacetime is
\begin{equation}
    H_{\rm vac}(r,\theta)
    =
    \frac{1}{2E_{\rm loc}(r,\theta)}
    \, U M^2 U^\dagger,
\end{equation}
with local energy given by
\begin{equation}
    E_{\rm loc}(r,\theta)
    = \sqrt{\frac{\Delta \Sigma}{A}}
      \left( E_\infty - \Omega_{\rm FD} L_z \right)
\end{equation}
for Kerr, and 
\begin{equation}
    E_{\rm loc}(r) = \sqrt{1 - \frac{2GM}{r}}\, E_\infty
\end{equation}
for Schwarzschild.

\paragraph{2. Spin--Curvature Coupling.}
From Sec.~\ref{sec:spin_curv},
\begin{equation}
    H_{\rm spin}
    =
    \frac{GM}{r^2}
    \left(
        \Sigma^{01} + \Sigma^{12} + \Sigma^{23}
    \right)
    +
    \frac{aGM}{r^{3}}
    \left(
        \Sigma^{03} + \cos\theta \, \Sigma^{13}
    \right)
    + \mathcal{O}(a^2).
\end{equation}

\paragraph{3. Frame-Dragging (Gravitomagnetic) Term.}
From Sec.~\ref{sec:kerr_framedrag},
\begin{equation}
    H_{\rm FD}
    =
    \Omega_{\rm FD}(r,\theta) \, \Sigma^{03}
    + \mathcal{O}(a^2),
\end{equation}
which in the equatorial plane reduces to
\begin{equation}
    H_{\rm FD}^{\rm eq}
    =
    \frac{2 G M a}{r^3}
    \, \Sigma^{03}.
\end{equation}

\paragraph{4. Gravitational Magnetic--Moment Term.}
From Sec.~3.7, the curvature-induced
magnetic-moment interaction is
\begin{equation}
    H_{\rm grav\text{-}mm}
    =
    \mu_G \frac{GM}{r^3}
    \left[
        \Sigma^{01}
        + \Sigma^{23}
        + a \cos\theta \, \Sigma^{03}
    \right]
    + \mathcal{O}(a^2).
\end{equation}
\paragraph{Limits and projection to the flavor sector.}
In the flat-space limit $GM/r \to 0$ and $a\to 0$, all curvature-dependent terms in Eq.~(56) vanish and the Hamiltonian reduces to the standard vacuum-plus-matter form
\begin{equation}
H_{\text{eff}} \;\longrightarrow\; \frac{1}{2E_\infty}\,U M^2 U^\dagger + H_{\text{matter}}\,,
\end{equation}
reproducing conventional three-flavor oscillations. In the ultrarelativistic regime $E_\infty \gg m_\nu$ considered here, helicity-flip transitions induced by spin couplings are suppressed by $m_\nu/E_\infty$ and are negligible for the energies relevant to our astrophysical benchmarks. Accordingly, we project the Hamiltonian onto the left-handed neutrino subspace and treat $H_{\text{spin}}$, $H_{\text{FD}}$, and $H_{\text{grav-mm}}$ as providing flavor-diagonal phase shifts and small corrections to effective mixing parameters rather than opening sizable spin-flip channels. This guarantees a smooth connection to the standard oscillation phenomenology while retaining the leading gravitational effects on flavor evolution.

\paragraph{5. Matter Potential.}
If neutrinos traverse an ambient medium with electron density $n_e$,
curved spacetime generalizes the MSW potential to
\begin{equation}
    H_{\rm matter}
    =
    \sqrt{2}G_F n_e(r) \,
    \mathrm{diag}(1,0,0),
\end{equation}
with $n_e(r)$ defined in the local rest frame.

\subsubsection*{Final Form}

Summing all contributions, the effective Hamiltonian in the flavor basis
is
\begin{equation}
\boxed{
\begin{aligned}
    H_{\rm eff}(r,\theta)
    &= \frac{1}{2E_{\rm loc}(r,\theta)} \, U M^2 U^\dagger 
    + \frac{GM}{r^2}
       \left( \Sigma^{01} + \Sigma^{12} + \Sigma^{23} \right)
\\
    &\quad + 
    \frac{aGM}{r^{3}}
        \left( \Sigma^{03} + \cos\theta \, \Sigma^{13} \right)
    + \Omega_{\rm FD}(r,\theta) \Sigma^{03}
\\
    &\quad + 
    \mu_G \frac{GM}{r^3}
        \left( \Sigma^{01} + \Sigma^{23} + a \cos\theta\, \Sigma^{03} \right)
    + \sqrt{2} G_F n_e(r) 
       \, {\rm diag}(1,0,0)
    + \mathcal{O}(a^2).
\end{aligned}}
\label{eq:Heff_final}
\end{equation}

This Hamiltonian contains all leading gravitational corrections relevant
for neutrino oscillations near compact objects, including redshift,
spin--curvature coupling, frame dragging, and geometric magnetic-moment
effects.

\subsubsection{Projection onto the Left-Handed Flavor Sector}

The operators $\Sigma^{ab}$ act on the spinor Hilbert space rather than
directly on flavor indices. Since astrophysical neutrinos satisfy
$E\gg m_i$, helicity-flip transitions are suppressed by $m_i/E$.
To leading order, the physical neutrino state is therefore confined to
the left-handed sector,

\[
|\nu_i\rangle \simeq |\nu_i,L\rangle .
\]

Projecting the spin-curvature Hamiltonian onto this subspace gives

\[
H^{(L)}_{\rm spin}
=
P_L H_{\rm spin} P_L ,
\qquad
P_L=\frac{1-\gamma^5}{2}.
\]

The expectation values
$\langle\Sigma^{ab}\rangle_L$
reduce the spin operators to c-number coefficients,
yielding an effective flavor-diagonal gravitational potential

\[
H^{(L)}_{\rm spin}
=
V_{\rm grav}(r,\theta)\,\mathbf 1_{\rm flavor}.
\]

The deterministic contribution therefore produces only a common phase,
while stochastic fluctuations of $V_{\rm grav}$ generate decoherence
through the Lindblad sector derived in Sec.~4.
%%%%%%%%%%%%%%%%%%%%%%%%%%%%%%%%%%%%%%%%%%%%%%%%%%%%%%%%%%%%%%%%%%%%%%%%
\section{Metric Fluctuations and Open-System Dynamics}
\label{sec:open}
%%%%%%%%%%%%%%%%%%%%%%%%%%%%%%%%%%%%%%%%%%%%%%%%%%%%%%%%%%%%%%%%%%%%%%%%

Neutrino propagation in curved spacetime is not strictly unitary when the underlying geometry contains stochastic or quantum fluctuations. Metric perturbations act as an environment that couples to the neutrino spinor field through the perturbed spin connection. In this section we develop a microscopic open-system description based on stationary spin-connection correlators, derive the Lindblad master equation in the Born–Markov–secular limit, and obtain a curvature-controlled decoherence rate and coherence length.

%%%%%%%%%%%%%%%%%%%%%%%%%%%%%%%%%%%%%%%%%%%%%%%%%%%%%%%%%%%%%%%%%%%%%%%%
\subsection{Metric fluctuations and perturbed spin connection}
\label{sec:metric-fluct}
%%%%%%%%%%%%%%%%%%%%%%%%%%%%%%%%%%%%%%%%%%%%%%%%%%%%%%%%%%%%%%%%%%%%%%%%

We model gravitational fluctuations by decomposing the metric into a smooth background $\bar g_{\mu\nu}$ and a perturbation $h_{\mu\nu}$,
\begin{equation}
g_{\mu\nu} = \bar g_{\mu\nu} + h_{\mu\nu},
\qquad
\langle h_{\mu\nu}(x) \rangle = 0,
\label{eq:g-split}
\end{equation}
where the average is taken with respect to a stationary gravitational state (which may describe quantized metric fluctuations, stochastic spacetime perturbations, or classical gravitational noise near compact objects).

Fluctuations of the tetrad and spin connection induce a perturbation of the Dirac covariant derivative. To first order in $h_{\mu\nu}$ we write
\begin{equation}
\Omega_\mu(x) = \bar\Omega_\mu(x) + \delta\Omega_\mu(x),
\qquad
\delta\Omega_\mu
=
\frac{1}{4}\,\delta\omega_{\mu}{}^{ab}\,\gamma_{[a}\gamma_{b]},
\label{eq:deltaOmega-def}
\end{equation}
where $\bar\Omega_\mu$ is the spin connection of the background metric $\bar g_{\mu\nu}$ and $\delta\omega_{\mu}{}^{ab}$ is linear in $h_{\mu\nu}$.

The perturbed Dirac equation then takes the form
\begin{equation}
i\gamma^\mu\big(\partial_\mu + \bar\Omega_\mu\big)\psi - m\psi
=
-\,i\gamma^\mu \delta\Omega_\mu\,\psi,
\label{eq:dirac-noise}
\end{equation}
where the right-hand side encodes the interaction between the neutrino spinor and metric fluctuations. This term will serve as the interaction Hamiltonian in the open-system treatment below.

%%%%%%%%%%%%%%%%%%%%%%%%%%%%%%%%%%%%%%%%%%%%%%%%%%%%%%%%%%%%%%%%%%%%%%%%
\subsection{Gaussian stationary noise and correlation functions}
\label{sec:noise-correlators}
%%%%%%%%%%%%%%%%%%%%%%%%%%%%%%%%%%%%%%%%%%%%%%%%%%%%%%%%%%%%%%%%%%%%%%%%

We treat the spin-connection perturbation $\delta\Omega_\mu(x)$ as a stationary, Gaussian noise process on the background spacetime. Its statistical properties are specified by the two-point correlator
\begin{equation}
\big\langle \delta\Omega_\mu(t,\mathbf{x})\,\delta\Omega_\nu(t-s,\mathbf{x}') \big\rangle
=
G_{\mu\nu}(s;\mathbf{x},\mathbf{x}'),
\label{eq:Gmunu-def}
\end{equation}
where stationarity implies that the correlator depends on the time difference $s$ only. For notational simplicity we suppress explicit spatial arguments and write $G_{\mu\nu}(s)$ when no confusion arises.

In a coarse-grained description appropriate for slowly varying neutrino wave packets, we approximate the correlator as local in space and characterized by a single correlation time $\tau_c(r)$ and amplitude matrix $\sigma_{\mu\nu}^2(r)$,
\begin{equation}
G_{\mu\nu}(s;r)
\simeq
\sigma_{\mu\nu}^2(r)\,e^{-|s|/\tau_c(r)}.
\label{eq:G-exp}
\end{equation}
Near compact objects we expect the fluctuation strength to be controlled by local curvature invariants. A natural choice is to relate $\sigma_{\mu\nu}^2$ to the Kretschmann scalar
\begin{equation}
K(r) \equiv R_{\alpha\beta\gamma\delta}R^{\alpha\beta\gamma\delta},
\end{equation}
by writing
\begin{equation}
\sigma_{\mu\nu}^2(r)
=
\alpha_{\mu\nu}\,\sqrt{K(r)},
\label{eq:sigma-K}
\end{equation}
with dimensionless coefficients $\alpha_{\mu\nu}=\mathcal{O}(1)$ encoding the tensor structure of the fluctuations and the dependence on the neutrino trajectory. In natural units ($c=\hbar=1$), $[K] = L^{-4}$ and $[\sqrt{K}] = L^{-2}$, so that $[G_{\mu\nu}] = L^{-2}$ as required.
%%%%%%%%%%%%%%%%%%%%%%%%%%%%%%%%%%%%%%%%%%%%%%%%%%%%%%%%%%

%%%%%%%%%%%%%%%%%%%%%%%%%%%%%%%%%%%%%%%%%%%%%%%%%%%%%%%%%%
\subsection{Microscopic Origin of Metric Fluctuations: Hawking Atmosphere}

To provide a concrete microscopic realization of the stochastic
gravitational environment introduced in the previous subsection,
we model the metric fluctuations using an effective Hawking-atmosphere
description surrounding a compact object. Our goal is not to claim that
Hawking radiation necessarily provides the dominant physical source of
astrophysical neutrino decoherence, but rather to employ a calculable
toy microscopic environment that permits a controlled derivation of
spin-connection correlation functions and Lindblad dynamics.

In realistic astrophysical settings, black holes are expected to possess
extremely small Hawking temperatures,
\begin{equation}
T_H = \frac{1}{8\pi GM},
\end{equation}
and therefore the direct physical impact of Hawking radiation on
high-energy neutrino propagation is likely to be negligible for
macroscopic black holes. Nevertheless, the Hawking-atmosphere framework
provides a theoretically well-defined stationary thermal environment
whose fluctuations can be used to illustrate how spacetime-induced
decoherence may emerge within an open-quantum-system treatment.

Within this effective description, fluctuations of the stress-energy
tensor source stochastic metric perturbations through the linearized
Einstein equations. The resulting spin-connection fluctuations then
induce nonunitary corrections to neutrino flavor evolution.
More generally, the formalism developed here may be interpreted as
parametrizing generic stationary spacetime fluctuations, with the
Hawking-atmosphere model serving primarily as an analytically tractable
benchmark realization.

The stochastic properties of the Hawking atmosphere are encoded
in the two-point correlation function

\begin{equation}
G_{\mu\nu}(s)
=
\Big\langle
\delta\Omega_\mu(t)
\delta\Omega_\nu(t-s)
\Big\rangle .
\end{equation}

For a stationary thermal environment the fluctuation-dissipation
theorem implies an approximately exponential decay,

\begin{equation}
G_{\mu\nu}(s)
=
A_{\mu\nu}
e^{-|s|/\tau_c},
\end{equation}

where $\tau_c$ denotes the Hawking-atmosphere correlation time.

The amplitude $A_{\mu\nu}$ can be estimated from the Einstein
equations. Since

\begin{equation}
h_{\mu\nu}
\sim
G\,\delta T_{\mu\nu},
\end{equation}

and thermal stress-energy fluctuations satisfy

\begin{equation}
\langle
\delta T\,\delta T
\rangle
\propto
T_H^8 ,
\end{equation}

one obtains

\begin{equation}
A_{\mu\nu}
\propto
G^2 T_H^8 .
\end{equation}

Using

\begin{equation}
T_H=\frac{1}{8\pi GM},
\end{equation}

gives

\begin{equation}
A_{\mu\nu}
\propto
\frac{1}{(GM)^6}.
\end{equation}
To express the result covariantly, it is convenient to rewrite
the fluctuation amplitude in terms of local curvature invariants.
For Schwarzschild spacetime,

\begin{equation}
K
=
R_{\mu\nu\rho\sigma}
R^{\mu\nu\rho\sigma}
=
\frac{48(GM)^2}{r^6}.
\end{equation}

The Hawking-atmosphere fluctuations therefore induce a
spin-connection correlator of the form

\begin{equation}
G_{\mu\nu}(s;r)
=
\alpha_{\mu\nu}
\sqrt{K(r)}
\,
e^{-|s|/\tau_c(r)},
\end{equation}

where $\alpha_{\mu\nu}$ is a dimensionless tensor determined by
the polarization structure of the thermal graviton bath.
%%%%%%%%%%%%%%%%%%%%%%%%%%%%%%%%%%%%%%%%%%%%%%%%%%%%%%%%%%%%%%%%%%%%%%%%
\subsection{Born--Markov--secular limit and Lindblad form}
\label{sec:born-markov}
%%%%%%%%%%%%%%%%%%%%%%%%%%%%%%%%%%%%%%%%%%%%%%%%%%%%%%%%%%%%%%%%%%%%%%%%

The interaction Hamiltonian associated with the fluctuating spin connection can be written as
\begin{equation}
H_{\text{int}}(t)
=
\int d^3x\,
\bar\psi(x)\,\gamma^\mu\,\delta\Omega_\mu(x)\,\psi(x),
\label{eq:Hint}
\end{equation}
where $x=(t,\mathbf{x})$ and $\bar\psi=\psi^\dagger\gamma^0$. Treating $\delta\Omega_\mu$ as an environment operator and working in the interaction picture, the total density matrix evolves as
\begin{equation}
\dot\rho_{\text{tot}}(t)
=
-\,i\,[H_{\text{int}}(t),\rho_{\text{tot}}(t)].
\end{equation}
Assuming an initially factorized state,
\begin{equation}
\rho_{\text{tot}}(t_0) = \rho_\nu(t_0)\otimes\rho_{\text{grav}},
\end{equation}
and applying the Born approximation (weak coupling, negligible backreaction on $\rho_{\text{grav}}$), we obtain the standard second-order master equation
\begin{equation}
\dot\rho_\nu(t)
=
-\,\int_0^\infty ds\,
\text{Tr}_{\text{grav}}
\Big\{
\big[H_{\text{int}}(t),\,
  [H_{\text{int}}(t-s),\rho_\nu(t)\otimes\rho_{\text{grav}}]
\big]
\Big\}.
\label{eq:born-master}
\end{equation}

Stationarity of the noise implies that all bath correlators depend only on $s$. In the Markov limit, where the correlation time $\tau_c(r)$ is much shorter than the characteristic system timescales (oscillation period, decoherence time), the upper limit of the $s$-integral can be extended to infinity and $\rho_\nu(t-s)$ replaced by $\rho_\nu(t)$. A secular (rotating-wave) approximation then diagonalizes the generator in the eigenbasis of the effective Hamiltonian $H_{\text{eff}}$ derived in Sec.~\ref{sec:ham}.

Under these conditions, the reduced density matrix obeys a Lindblad master equation of the form
\begin{equation}
\dot\rho_\nu
=
-\,i\,[H_{\text{eff}},\rho_\nu]
+
\frac{1}{2}
\sum_i
\left(
L_i\,\rho_\nu\,L_i^\dagger
-
\frac{1}{2}\{L_i^\dagger L_i,\rho_\nu\}
\right),
\label{eq:lindblad-general}
\end{equation}
where the Lindblad operators $L_i$ arise from the spectral decomposition of the integrated correlation functions $G_{\mu\nu}(s)$ and encode the coupling of spin degrees of freedom to curvature-induced noise. For curvature-dominated regimes, dimensional analysis and explicit evaluation of the spin-connection fluctuations lead to operators of the schematic form
For curvature-induced spin-connection fluctuations, the
dissipative dynamics can be written in Lindblad form,

\begin{equation}
\dot{\rho}_\nu
=
-i[H_{\rm eff},\rho_\nu]
+
\sum_i
\left(
L_i\rho_\nu L_i^\dagger
-
\frac12
\{L_i^\dagger L_i,\rho_\nu\}
\right).
\label{eq:LindbladGeneral}
\end{equation}

The Lindblad operators are obtained by diagonalizing the
Kossakowski matrix constructed from the spin-connection
correlation functions. Explicitly,

\begin{equation}
L_i
=
\sqrt{\lambda_i}\,
\Sigma_i ,
\label{eq:LindbladOperators}
\end{equation}

where $\Sigma_i$ denote the relevant spin generators and
$\lambda_i$ are the non-negative eigenvalues of the
Kossakowski matrix,

\begin{equation}
C_{ij}
=
\int_0^\infty ds\,
\langle
B_i(s)B_j(0)
\rangle .
\label{eq:KossakowskiMain}
\end{equation}

For the Hawking-atmosphere correlator introduced in
Sec.~4.2, one finds

\begin{equation}
\lambda_i
=
\alpha_i\,
\tau_c(r)\,
\sqrt{K(r)},
\label{eq:EigenvaluesMain}
\end{equation}

with $\alpha_i\ge0$. The positivity of the eigenvalues
guarantees that the Kossakowski matrix is positive
semidefinite and therefore that the resulting evolution is
completely positive and trace preserving. The full
construction is presented in Appendix~\ref{app:cp}. The Lindblad operators are obtained from the spectral
decomposition of the Kossakowski matrix associated with the
spin-connection correlators,

\begin{equation}
L_i=\sqrt{\lambda_i}\,\Sigma_i ,
\end{equation}

where $\lambda_i$ are the non-negative eigenvalues of
$C_{ij}$. The explicit diagonalization and proof of complete
positivity are given in Appendix~\ref{app:cp}.
\subsection{Microscopic Graviton-Bath Derivation of the Curvature-Induced Decoherence Rate}\label{sec}

A central ingredient of the present framework is the gravitational decoherence rate $\Gamma_{\rm grav}$ governing the damping of off-diagonal flavor coherences in the reduced neutrino density matrix. While the phenomenological analysis developed in the main text employs a curvature-dependent decoherence coefficient, it is desirable to establish a microscopic origin for the corresponding Lindblad term. In this section we derive the decoherence rate from stochastic spin-connection fluctuations generated by a graviton environment propagating on a curved spacetime background.

\subsubsection{Linearized Gravitational Environment}

We decompose the spacetime metric into a classical background and a fluctuating component,

\begin{equation} \bar g_{\mu\nu}(x)+\kappa h_{\mu\nu}(x),\qquad\kappa=\sqrt{32\pi G},\end{equation}

where $\bar g_{\mu\nu}$ denotes the Schwarzschild or Kerr background metric and $h_{\mu\nu}$ represents quantum gravitational fluctuations.

The tetrad field similarly decomposes as

\begin{equation}
e^a_{\mu}
=
\bar e^a_{\mu}
+
\delta e^a_{\mu}.
\end{equation}

with

\begin{equation}
\delta e^a_{\mu}
=
\frac{\kappa}{2}
h_{\mu\nu}
\bar e^{a\nu}.
\end{equation}

The spin connection therefore becomes

\begin{equation}
\Omega_\mu
=
\bar\Omega_\mu
+
\delta\Omega_\mu .
\end{equation}
where

\begin{equation}
\delta\Omega_\mu
=
\frac14
\delta\omega_{\mu}^{ab}
\gamma_{[a}\gamma_{b]} .
\end{equation}

Since

\begin{equation}\delta\omega\sim\partial h,\end{equation}

spin-connection fluctuations are induced directly by graviton fluctuations.

\subsubsection{Neutrino--Graviton Interaction Hamiltonian}

Starting from the covariant Dirac equation

\begin{equation}\left(i\gamma^\mu D_\mu-m\right)\psi=0,\end{equation}

the interaction Hamiltonian generated by spin-connection fluctuations is
\begin{equation}
\int d^3x,\bar\psi(x)\gamma^\mu\delta\Omega_\mu(x)\psi(x).\end{equation}

Projecting onto the ultra-relativistic neutrino sector yields

\begin{equation}
H_{\rm int}
=
\sum_i S_i \otimes B_i ,
\end{equation}

where \(S_i = \frac{1}{2}\Sigma_i\) act on the neutrino Hilbert space, while
\(B_i \propto \delta\Omega_i\) represent environmental operators.

The total Hamiltonian therefore assumes the standard open-system form
\begin{equation}
H_{\rm tot}
=
H_\nu + H_{\rm grav} + H_{\rm int}.
\end{equation}

\subsubsection{Graviton Correlation Functions and Spectral Density}

The environmental fluctuations are characterized by the two-point correlation function
\begin{equation}
C_{ij}(s)
=
\langle B_i(s) B_j(0) \rangle .
\end{equation}

Equivalently, the correlator may be expressed through a graviton spectral density,
\begin{equation}
J_{ij}(\omega)
=
\int_{-\infty}^{\infty} ds\,
e^{i\omega s}
\langle B_i(s) B_j(0) \rangle .
\end{equation}

\begin{equation}
C_{ij}(s)
=
\int_{0}^{\infty} d\omega \,
J_{ij}(\omega)\,
e^{-i\omega s},
\end{equation}
where \(J_{ij}(\omega)\) encodes the spectrum of stochastic spin-connection fluctuations induced by the gravitational environment.
The corresponding Kossakowski matrix is
\begin{equation}
C_{ij}
=
\int_{0}^{\infty} d\omega \,
J_{ij}(\omega).
\end{equation}

Diagonalizing \(C_{ij}\) yields eigenvalues \(\lambda_i\), and the Lindblad operators take the form
\begin{equation}
L_i
=
\sqrt{\lambda_i}\,\Sigma_i .
\end{equation}
Thus the decoherence coefficients are determined directly by the graviton spectral density.

\subsubsection{Born--Markov Master Equation}

Assuming weak coupling between the neutrino sector and the gravitational environment, the total density matrix factorizes as
\begin{equation}
\rho_{\rm tot}
\simeq
\rho_\nu \otimes \rho_{\rm grav}.
\end{equation}

The Born--Markov approximation then yields the reduced evolution equation
\begin{align}
\frac{d\rho_\nu(t)}{dt}
=
-\int_0^\infty ds \,
{\rm Tr}_{\rm grav}
\Big[
H_{\rm int}(t),
\big[
H_{\rm int}(t-s),
\rho_\nu(t)\otimes\rho_{\rm grav}
\big]
\Big].
\end{align}
After performing the secular approximation, one obtains the Lindblad equation
\begin{equation}
\frac{d\rho_\nu}{dt}
=
-i\left[H_{\rm eff},\rho_\nu\right]
+
\sum_i \lambda_i
\left(
L_i \rho_\nu L_i^\dagger
-
\frac{1}{2}
\left\{
L_i^\dagger L_i,
\rho_\nu
\right\}
\right).
\end{equation}

\subsubsection{Curvature Scaling of the Correlator}

The spin connection is constructed from first derivatives of the tetrad field,
\begin{equation}
\Omega_\mu^{\;\;ab}
=
\mathcal{O}(\partial e),
\end{equation}
whereas the Riemann curvature tensor involves one additional derivative,
\begin{equation}
R_{\mu\nu\rho\sigma}
=
\mathcal{O}(\partial \Omega)
=
\mathcal{O}(\partial^2 e).
\end{equation}

For vacuum Schwarzschild and Kerr spacetimes, the natural local curvature scale is provided by the Kretschmann invariant,
\begin{equation}
K
=
R_{\mu\nu\rho\sigma}R^{\mu\nu\rho\sigma}.
\end{equation}

Because the spin connection has dimensions of inverse length,
\begin{equation}
[\Omega_\mu]=L^{-1},
\end{equation}
while
\begin{equation}
[K]=L^{-4},
\end{equation}
the unique local quantity with the correct dimensions is
\begin{equation}
\delta\Omega
\sim
K^{1/4}.
\end{equation}
Consequently,
\begin{equation}
\langle \delta\Omega(s)\,\delta\Omega(0) \rangle
\sim
K^{1/2}.
\end{equation}

Assuming stationary Gaussian fluctuations, the corresponding correlation function may be parametrized as
\begin{equation}
G(s)
=
\alpha \sqrt{K}\,
e^{-s/\tau_c},
\end{equation}
\begin{equation}
G(s)
\equiv
\langle \delta\Omega(s)\,\delta\Omega(0) \rangle
=
\alpha \sqrt{K}\,
e^{-s/\tau_c}.
\end{equation}

The parametrization
\[
G(s)=\alpha\sqrt{K}\,e^{-s/\tau_c}
\]
should be interpreted as an effective low-energy closure relation for the
spin-connection correlator. A complete determination of the coefficient
\(\alpha\) and the exact curvature dependence would require evaluation of
the graviton two-point function in the Schwarzschild or Kerr background,
which lies beyond the scope of the present work. The present approach
therefore combines a microscopic open-system derivation with an EFT
parameterization of the gravitational noise kernel.
\subsubsection{Evaluation of the Lindblad Coefficient}

Substituting the correlation function into the Kossakowski integral yields
\begin{align}
\Gamma_{\rm grav}
&=
\int_0^\infty ds\, G(s)
\nonumber\\
&=
\alpha \sqrt{K}
\int_0^\infty ds\, e^{-s/\tau_c}
\nonumber\\
&=
\alpha \tau_c \sqrt{K}.
\end{align}

Therefore,
\begin{equation}
\Gamma_{\rm grav}
=
\alpha \tau_c \sqrt{K}.
\end{equation}

This expression provides a microscopic justification for the curvature-dependent decoherence rate employed throughout this work.

\subsubsection{Why \texorpdfstring{$\sqrt{K}$}{square-root K}}

For vacuum Schwarzschild and Kerr geometries,
\begin{equation}
R = 0,
\qquad
R_{\mu\nu} = 0,
\end{equation}
which immediately implies
\begin{equation}
R_{\mu\nu}R^{\mu\nu} = 0.
\end{equation}

The leading nonvanishing scalar curvature invariant is therefore the Kretschmann scalar,
\begin{equation}
K
=
R_{\mu\nu\rho\sigma}
R^{\mu\nu\rho\sigma}.
\end{equation}
For vacuum Schwarzschild and Kerr geometries,
\begin{equation}
R=0,
\qquad
R_{\mu\nu}=0,
\end{equation}
which immediately implies
\begin{equation}
R_{\mu\nu}R^{\mu\nu}=0.
\end{equation}

The leading nonvanishing scalar curvature invariant is therefore
\begin{equation}
K
=
R_{\mu\nu\rho\sigma}
R^{\mu\nu\rho\sigma}.
\end{equation}

The spin connection carries dimensions
\begin{equation}
[\Omega_\mu]=L^{-1},
\end{equation}
whereas
\begin{equation}
[K]=L^{-4}.
\end{equation}

Consequently, the unique local curvature quantity possessing the
correct dimensions is \(K^{1/4}\). We therefore parametrize the
root-mean-square spin-connection fluctuation amplitude as
\begin{equation}
\delta\Omega_{\rm rms}
=
\xi K^{1/4},
\end{equation}
where \(\xi\) is a dimensionless coefficient encoding the
microscopic properties of the gravitational environment.

Since the Lindblad coefficient originates from a two-point correlator
of spin-connection fluctuations and each fluctuation scales as
\begin{equation}
\delta\Omega \sim \delta\Omega_{\rm rms}
\sim \xi K^{1/4},
\end{equation}
the corresponding correlator scales as
\begin{equation}
\langle \delta\Omega(s)\delta\Omega(0)\rangle
\sim
K^{1/2}.
\end{equation}
Since the Lindblad coefficient originates from a two-point correlator
of spin-connection fluctuations and each fluctuation scales as
\begin{equation}
\delta\Omega \sim K^{1/4},
\end{equation}
the corresponding correlator scales as
\begin{equation}
\langle \delta\Omega(s)\delta\Omega(0)\rangle
\sim
K^{1/2}.
\end{equation}

The Born--Markov integral therefore yields
\begin{equation}
\Gamma_{\rm grav}
\propto
\sqrt{K}.
\end{equation}

\subsubsection{Schwarzschild and Kerr Limits}

For Schwarzschild spacetime, the Kretschmann scalar is
\begin{equation}
K_{\rm Schw}
=
\frac{48(GM)^2}{r^6},
\end{equation}
which gives
\begin{equation}
\Gamma_{\rm grav}^{\rm Schw}
=
\alpha \tau_c
\sqrt{\frac{48(GM)^2}{r^6}}
=
4\sqrt{3}\,
\alpha\tau_c
\frac{GM}{r^3}.
\end{equation}

For Kerr spacetime,
\begin{equation}
\Gamma_{\rm grav}^{\rm Kerr}
=
\alpha\tau_c
\sqrt{K_{\rm Kerr}},
\end{equation}
where \(K_{\rm Kerr}\) denotes the Kerr Kretschmann scalar. This expression demonstrates explicitly how black-hole rotation modifies the decoherence rate through the local spacetime curvature.
%%%%%%%%%%%%%%%%%%%%%%%%%%%%%%%%%%%%%%%%%%%%%%%%%%%%%%%%%%%%%%%%%%%%%%%%
\subsection{Curvature-controlled decoherence rate and correlation time}
\label{sec:Gamma-curv}
%%%%%%%%%%%%%%%%%%%%%%%%%%%%%%%%%%%%%%%%%%%%%%%%%%%%%%%%%%%%%%%%%%%%%%%%
The decoherence rate governing the nonunitary evolution of the neutrino
density matrix is determined by the integrated spin-connection
correlation function generated by the Hawking atmosphere. Within the
Born--Markov approximation, the dissipative coefficient appearing in the
Lindblad equation is obtained from the time integral of the stationary
correlator,

\begin{equation}
\Gamma_{\rm grav}(r)
=
\int_{0}^{\infty} ds\,
G(s;r),
\label{eq:GammaDef}
\end{equation}

where $G(s;r)$ denotes the effective spin-connection two-point
correlation function introduced in the previous subsection. Substituting
the Hawking-atmosphere correlator

\begin{equation}
G_{\mu\nu}(s;r)
=
\alpha_{\mu\nu}
\sqrt{K(r)}
\,e^{-|s|/\tau_c(r)},
\label{eq:HawkingCorrelator}
\end{equation}

and performing the integration yields

\begin{equation}
\Gamma_{\rm grav}(r)
=
\alpha\,\tau_c(r)\sqrt{K(r)},
\label{eq:GammaGeneral}
\end{equation}

where $\alpha$ is an effective dimensionless coefficient that absorbs the
tensor structure of the correlator and numerical factors arising from the
projection onto the neutrino flavor sector. Equation
(\ref{eq:GammaGeneral}) demonstrates that the strength of decoherence is
controlled jointly by the local spacetime curvature and the microscopic
correlation time of the Hawking atmosphere.

For Schwarzschild spacetime, the relevant curvature invariant is the
Kretschmann scalar

\begin{equation}
K(r)
=
R_{\mu\nu\rho\sigma}R^{\mu\nu\rho\sigma}
=
\frac{48(GM)^2}{r^6},
\label{eq:Kretschmann}
\end{equation}

which gives

\begin{equation}
\Gamma_{\rm grav}(r)
=
\alpha\sqrt{48}\,
\tau_c(r)\,
\frac{GM}{r^3}.
\label{eq:GammaSchwarzschild}
\end{equation}

The decoherence rate therefore increases rapidly as the neutrino
approaches a compact object, reflecting the growth of local curvature
and the enhanced interaction with Hawking-atmosphere fluctuations. This
result provides a microscopic origin for the curvature-dependent
decoherence scale employed throughout the remainder of this work.

The corresponding coherence length is defined as

\begin{equation}
L_{\rm coh}(r)
\equiv
\Gamma_{\rm grav}^{-1}(r).
\label{eq:LcohDef}
\end{equation}

If the correlation time is approximately constant over the propagation
region, Eq.~(\ref{eq:GammaSchwarzschild}) implies

\begin{equation}
\Gamma_{\rm grav}(r)
\propto
\frac{GM}{r^3},
\qquad
L_{\rm coh}(r)
\propto
\frac{r^3}{GM},
\label{eq:GeometricScaling}
\end{equation}

recovering the geometric scaling relation emphasized in this work.
Alternatively, if $\tau_c(r)$ is determined by a local curvature scale,
additional corrections arise through its radial dependence, but the
dominant curvature dependence remains governed by the Kretschmann
invariant.
where $\alpha$ is an effective dimensionless coefficient collecting tensor indices and numerical factors. In natural units, $[\sqrt{K}] = L^{-2}$ and $[\tau_c]=L$, so that $[\Gamma_{\text{grav}}]=L^{-1}$ as required for a rate.

For Schwarzschild spacetime,
\begin{equation}
K(r)
=
\frac{48(GM)^2}{r^6},
\qquad
\sqrt{K(r)} = \frac{\sqrt{48}\,GM}{r^3},
\end{equation}
so that
\begin{equation}
\Gamma_{\text{grav}}(r)
=
\gamma_0\,\tau_c(r)\,\frac{GM}{r^3},
\label{eq:Gamma-Sch-tau}
\end{equation}
with $\gamma_0=\mathcal{O}(1)$ absorbing numerical factors and the detailed tensor structure of $\alpha_{\mu\nu}$.

To make contact with the simple geometric scaling used in the phenomenological sections, it is convenient to express $\tau_c(r)$ in terms of a local curvature length scale. A natural choice is to take
\begin{equation}
\ell_c(r)
\sim
|R_{\mu\nu\rho\sigma}|^{-1/2}
\sim
\left(\frac{r^3}{GM}\right)^{1/2},
\end{equation}
and to parametrize
\begin{equation}
\tau_c(r) = \kappa\,\ell_c(r),
\end{equation}
with $\kappa$ a dimensionless constant encoding details of the gravitational environment. In that case,
\begin{equation}
\Gamma_{\text{grav}}(r)
\sim
\gamma_0\,\kappa\,\frac{GM}{r^3}\,
\left(\frac{r^3}{GM}\right)^{1/2}
=
(\gamma_0\kappa)\,\frac{\sqrt{GM}}{r^{3/2}}\,,
\end{equation}
and the associated coherence length behaves as
\begin{equation}
L_{\text{coh}}(r)
\equiv
\Gamma_{\text{grav}}^{-1}(r)
\sim
\frac{r^{3/2}}{\sqrt{GM}}.
\end{equation}
Alternatively, for environments where $\tau_c(r)$ is set by external matter or turbulence scales rather than pure curvature, it is often convenient to treat $\tau_c$ as approximately constant over the region of interest. In that case Eq.~\eqref{eq:Gamma-Sch-tau} gives
\begin{equation}
\Gamma_{\text{grav}}(r)
\propto
\frac{GM}{r^3},
\qquad
L_{\text{coh}}(r)
\propto
\frac{r^3}{GM},
\label{eq:Lcoh-simple}
\end{equation}
up to an overall factor of $\tau_c$ and dimensionless coefficients. It is this purely geometric scaling that we emphasize in the main text, while keeping the explicit $\tau_c$ dependence in our quantitative estimates in Secs.~5 and 7.

%%%%%%%%%%%%%%%%%%%%%%%%%%%%%%%%%%%%%%%%%%%%%%%%%%%%%%%%%%%%%%%%%%%%%%%%
\subsection{Final master equation and coherence length}
\label{sec:master-final}
%%%%%%%%%%%%%%%%%%%%%%%%%%%%%%%%%%%%%%%%%%%%%%%%%%%%%%%%%%%%%%%%%%%%%%%%

Combining the unitary dynamics generated by the effective Hamiltonian $H_{\text{eff}}(r,\theta)$ in Eq.~(56) with the curvature-controlled dissipative term characterized by $\Gamma_{\text{grav}}(r)$, the evolution of the reduced neutrino density matrix can be written, in the flavor basis, as
\begin{equation}
\dot\rho_\nu(t)
=
-\,i\,[H_{\text{eff}}(r,\theta),\rho_\nu(t)]
-
\Gamma_{\text{grav}}(r)\,
\big(\rho_\nu(t) - \text{diag}[\rho_\nu(t)]\big),
\label{eq:master-final}
\end{equation}
where the dissipator has been written in a simplified diagonal form appropriate for the flavor basis and the dominant Lindblad operator structure in Eq.~(83). Equation~\eqref{eq:master-final} implies exponential damping of off-diagonal flavor components,
\begin{equation}
\rho_{\alpha\beta}(t)
=
\rho_{\alpha\beta}(0)\,\exp\big[-\Gamma_{\text{grav}}(r)\,t\big],
\qquad
\alpha\neq\beta.
\end{equation}

The associated coherence length is defined as
\begin{equation}
L_{\text{coh}}(r)
\equiv
\Gamma_{\text{grav}}^{-1}(r),
\end{equation}
and, for Schwarzschild backgrounds, exhibits the geometric scaling in Eq.~\eqref{eq:Lcoh-simple}, up to the choice of correlation time $\tau_c(r)$. We emphasize that this scaling is controlled by local curvature invariants and therefore provides a direct geometric link between spacetime curvature and the loss of neutrino flavor coherence near compact astrophysical objects. In Sec.5 we incorporate $\Gamma_{\text{grav}}(r)$ into the oscillation probabilities, and in Sec.7 we quantify its impact on astrophysical flavor ratios and event rates for representative choices of $\tau_c$.

\section{Oscillation Probabilities in Curved Spacetime}
\label{sec:osc_prob}

Neutrino flavor transition probabilities follow from the solution of the
open-system evolution equation
\begin{equation}
    \dot{\rho}
    =
    - i \left[ H_{\rm eff}(r,\theta),\, \rho \right]
    - \Gamma_{\rm grav}(r)
      \left( \rho - \mathrm{diag}(\rho) \right),
    \label{eq:master_osc}
\end{equation}
where $H_{\rm eff}$ is the Hamiltonian derived in
Sec.~\ref{sec:combined_H}. Below we extract the corresponding
oscillation probabilities.

\subsection{Redshift--Corrected Oscillation Phase}

For two flavors $(\nu_\alpha \leftrightarrow \nu_\beta)$ the unitary
portion of the evolution generates a gravitationally modified phase
\begin{equation}
    \phi_{ij}(r,\theta)
    =
    \int^{t}_{0} dt'\;
    \frac{\Delta m^2_{ij}}{2 E_{\rm loc}(r(t'),\theta(t'))}.
    \label{eq:redshift_phase}
\end{equation}

For radial propagation in Schwarzschild spacetime,
\begin{equation}
    \phi_{ij}^{\rm Schw}(r)
    =
    \int \frac{\Delta m^2_{ij}}{2E_\infty}
    \left( 1 - \frac{2GM}{r} \right)^{-1/2} dt.
\end{equation}

In Kerr spacetime, substituting Eq.~\eqref{eq:kerr_loc_energy} gives
\begin{equation}
    \phi_{ij}^{\rm Kerr}(r,\theta)
    =
    \int dt \;
    \frac{\Delta m^2_{ij}}
         {2\sqrt{\tfrac{\Delta\Sigma}{A}}\,
          \big(E_\infty - \Omega_{\rm FD} L_z\big)}.
\end{equation}

\subsection{Frame--Dragging Corrections}

Frame dragging produces an additional spin-induced phase contribution,
\begin{equation}
    \phi_{\rm FD}(r,\theta)
    =
    \int dt \;
    \Omega_{\rm FD}(r,\theta)\,
    \bra{\nu}\Sigma^{03}\ket{\nu}.
\end{equation}

For equatorial motion $(\theta = \pi/2)$,
\begin{equation}
    \phi_{\rm FD}^{\rm eq}
    =
    \int dt \;
    \frac{2GMa}{r^3} \, S_z,
\end{equation}
with $S_z$ the spin projection along the rotation axis.

\subsection{Oscillation Probability with Curvature Decoherence}

Solving Eq.~\eqref{eq:master_osc} yields the flavor transition
probability
\begin{equation}
\boxed{
\begin{aligned}
P_{\alpha\rightarrow\beta}(L)
&=
\delta_{\alpha\beta}
- 4
  \sum_{i>j}
  \Re\!\left(
     U_{\alpha i}U_{\beta i}^\ast
     U_{\alpha j}^\ast U_{\beta j}
  \right)
  \exp\!\left[-\Gamma_{\rm grav}(L)\, L\right]
  \sin^2\!\left(
      \frac{\phi_{ij}(L) + \phi^{\rm FD}(L)}{2}
  \right)
\\[6pt]
&\quad
+ 2
  \sum_{i>j}
  \Im\!\left(
     U_{\alpha i}U_{\beta i}^\ast
     U_{\alpha j}^\ast U_{\beta j}
  \right)
  \exp\!\left[-\Gamma_{\rm grav}(L)\, L\right]
  \sin\!\left(
      \phi_{ij}(L) + \phi^{\rm FD}(L)
  \right).
\end{aligned}}
\end{equation}

The exponential damping term suppresses the interference responsible for
oscillations.
\subsection{Analytical Structure of the Curvature-Corrected Probability}

Equation~(84) contains three distinct gravitational contributions
to neutrino flavor evolution:

\begin{enumerate}

\item {\bf Gravitational redshift modification.}

The oscillation phase depends on the locally measured energy,
\begin{equation}
E_{\rm loc}(r,\theta)
=
\sqrt{-g_{tt}}\,E_\infty ,
\end{equation}
which modifies the vacuum oscillation length according to
\begin{equation}
L_{\rm osc}^{\rm grav}
=
\frac{4\pi E_{\rm loc}}{\Delta m^2}.
\end{equation}

Near compact objects, the reduction of $E_{\rm loc}$ enhances
the effective oscillation phase and shifts the interference pattern.

\item {\bf Kerr frame dragging.}

In rotating geometries, the additional phase
\begin{equation}
\phi_{\rm FD}
=
\int dt\,\Omega_{\rm FD}(r,\theta)\,
\langle \Sigma^{03}\rangle
\end{equation}
introduces a gravitomagnetic correction to flavor evolution.

This contribution vanishes in the Schwarzschild limit
$(a\to0)$ and becomes maximal near rapidly rotating black holes.

\item {\bf Curvature-induced decoherence.}

The exponential factor
\begin{equation}
\exp[-\Gamma_{\rm grav}(r)L]
\end{equation}
suppresses off-diagonal density-matrix elements through
entanglement with gravitational fluctuations.

Using
\begin{equation}
\Gamma_{\rm grav}(r)
\sim
\frac{GM}{r^3},
\end{equation}
the coherence length becomes
\begin{equation}
L_{\rm coh}(r)
\sim
\frac{r^3}{GM},
\end{equation}
demonstrating that coherence is strongly reduced near
the horizon and restored asymptotically at large distances.

\end{enumerate}

The interplay between these three effects produces a
nontrivial transition between coherent oscillations and
gravitationally induced flavor decoherence.
In particular, low-energy neutrinos accumulate larger
gravitational phases and therefore exhibit stronger
modifications near compact objects.
\subsection{Numerical Framework and Benchmark Parameters}
\label{sec:numericalsetup}

Throughout the numerical analysis we adopt the current global best-fit
three-flavor oscillation parameters,
\begin{align}
\Delta m_{21}^{2} &= 7.42 \times 10^{-5}\ \mathrm{eV}^{2}, \\
\Delta m_{31}^{2} &= 2.517 \times 10^{-3}\ \mathrm{eV}^{2}, \\
\sin^{2}\theta_{12} &= 0.304, \\
\sin^{2}\theta_{23} &= 0.573, \\
\sin^{2}\theta_{13} &= 0.02219, \\
\delta_{\mathrm{CP}} &= 197^{\circ},
\end{align}
consistent with recent global analyses~\cite{Esteban2024,NuFIT2024}.

The benchmark compact-object configurations used throughout this work are
summarized in Table~\ref{tab:benchmarks}.

\begin{table}[h]
\centering
\caption{Benchmark compact-object parameters.}
\label{tab:benchmarks}
\begin{tabular}{cccc}
\hline
Benchmark & Mass & Spin & Geometry \\
\hline
B1 & $10\,M_{\odot}$ & 0 & Schwarzschild \\
B2 & $10\,M_{\odot}$ & 0.5 & Moderate Kerr \\
B3 & $10\,M_{\odot}$ & 0.95 & Near-extremal Kerr \\
\hline
\end{tabular}
\end{table}

The neutrino energies considered are
\begin{equation}
E_{\nu} = \{5,\; 20,\; 100,\; 500\}\ \mathrm{GeV},
\end{equation}
representative of the energy range relevant for atmospheric,
astrophysical, and next-generation neutrino observatories.

For the curvature-induced decoherence sector we define
\begin{equation}
\Gamma_{\mathrm{grav}} = \alpha \tau_{c} \sqrt{K},
\end{equation}
and adopt the benchmark values
\begin{equation}
\alpha = \left\{ 10^{-4},\; 10^{-3},\; 10^{-2},\; 10^{-1},\; 1 \right\},
\end{equation}
together with
\begin{equation}
\tau_{c} = \left\{ 10^{-6},\; 10^{-5},\; 10^{-4},\; 10^{-3} \right\}\ \mathrm{s}.
\end{equation}
These values span the transition between effectively coherent
and strongly decohering propagation regimes. 
\begin{table*}[t]
\centering
\caption{
Detector characteristics adopted in the forecast sensitivity analysis. 
The effective volumes, energy ranges, flavor-identification capabilities, 
and projected exposures are representative of the design performance of 
IceCube-Gen2, KM3NeT, and P-ONE \cite{IceCubeGen2WhitePaper,KM3NeTLoI,PONE2023}. 
These benchmarks are used to estimate the sensitivity of future high-energy 
neutrino observatories to curvature-induced decoherence and gravitational 
flavor-memory effects.
}
\label{tab:detector_response}
\begin{tabular}{lccccc}
\hline
Detector & Effective Volume & Energy Range & Flavor Resolution & Exposure & Reference \\
\hline
IceCube-Gen2 & $\sim 8~{\rm km}^{3}$ & $10^{2}$--$10^{7}~{\rm GeV}$ & Track/Cascade & 10 yr & Ref.~\cite{IceCubeGen2WhitePaper} \\
KM3NeT       & $\sim 1~{\rm km}^{3}$ & $10^{2}$--$10^{6}~{\rm GeV}$ & Track/Cascade & 10 yr & Ref.~\cite{KM3NeTLoI} \\
P-ONE        & $\sim 1~{\rm km}^{3}$ & $10^{2}$--$10^{6}~{\rm GeV}$ & Track/Cascade & 10 yr & Ref.~\cite{PONE2023} \\
\hline
\end{tabular}
\end{table*}
The detector benchmarks summarized in Table~\ref{tab:detector_response}
serve as the basis for all projected sensitivity estimates presented in
this work. For each detector, the expected event yields are computed from
the flavor-dependent neutrino fluxes obtained after propagation through
the curved spacetime background, including the effects of gravitational
redshift, spin-curvature coupling, frame dragging, and curvature-induced
decoherence. The resulting flavor compositions are subsequently folded
with the detector response assumptions listed in
Table~\ref{tab:detector_response} to construct likelihood functions and
projected exclusion contours in the
$(\alpha,\tau_c)$ parameter space. \subsection{Likelihood Construction}

To quantify the sensitivity of future neutrino observatories to 
curvature-induced decoherence, we define
\begin{equation}
\chi^{2} = \sum_i \frac{\left[ N_i^{\rm model} - N_i^{\rm SM} \right]^2}{\sigma_i^2},
\end{equation}
where
\begin{equation}
\sigma_i^2 = N_i^{\rm SM} + \left(f_{\rm sys}\, N_i^{\rm SM}\right)^2.
\end{equation}
The likelihood is
\begin{equation}
\mathcal{L} \propto e^{-\chi^2/2}.
\end{equation}
Projected exclusion contours correspond to
\begin{equation}
\Delta\chi^2 = 2.30, \quad 5.99, \quad 9.21,
\end{equation}
representing 68\%, 95\%, and 99\% confidence regions, respectively.
To illustrate these effects quantitatively, we now present
a numerical analysis of the survival probability in both
Schwarzschild and Kerr geometries.
\begin{figure*}[htp]

\centering
\includegraphics[width=0.98\textwidth]{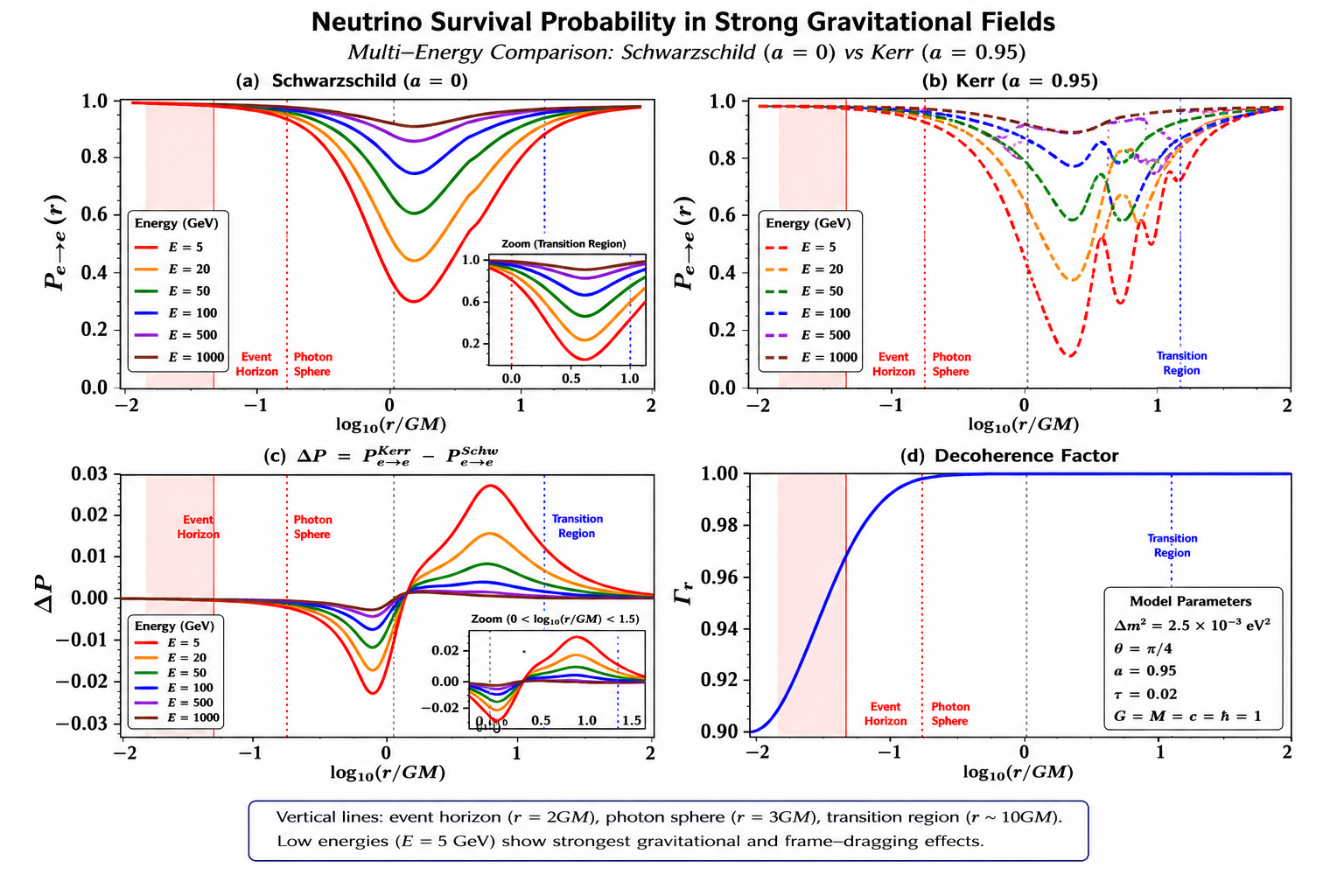}

\caption{Multi-panel illustration of neutrino oscillations in strong
gravitational fields for Schwarzschild and Kerr geometries.(a) Survival probability in Schwarzschild spacetime showing curvature-induced suppression of flavor coherence. (b) Kerr spacetime results including frame-dragging effects, which induce additional oscillatory phase shifts. (c) Difference $\Delta P \equiv P^{\rm Kerr}_{e\to e}-P^{\rm Schw}_{e\to e}$,
isolating the pure rotational contribution of the black hole. (d) Curvature-induced decoherence factor $e^{-\Gamma_{\rm grav} r}$, demonstrating suppression of quantum coherence near the horizon. Vertical lines indicate the event horizon $(r=2GM)$, photon sphere $(r=3GM)$, and transition region between strong- and weak-gravity regimes. The figure is generated using the oscillation probability derived in Eq.~(84), including gravitational redshift, frame dragging, and Lindblad decoherence effects.}

\label{fig:gravity_oscillations}

\end{figure*}
The behavior shown in Fig.~\ref{fig:gravity_oscillations}
demonstrates the interplay between gravitational redshift,
frame dragging, and curvature-induced decoherence.

Near the horizon, strong curvature suppresses quantum
coherence through the exponential damping factor
$e^{-\Gamma_{\rm grav}L}$,
leading to reduced oscillation amplitudes.
In Kerr spacetime, additional phase shifts generated by
frame dragging produce visible deviations from the
Schwarzschild case, particularly in the transition region
between strong and weak gravity.

The effect is strongest for lower-energy neutrinos,
which accumulate larger gravitational phases during propagation.
\subsection{Gravitational Flavor Memory Observable}
The local oscillation probability provides only an instantaneous
measure of flavor evolution.
However, in curved spacetime, neutrinos accumulate nontrivial
phase distortions over extended propagation distances due to
gravitational redshift, frame dragging, and curvature-induced
decoherence.

To quantify the cumulative gravitational imprint on flavor
propagation, we introduce a gravitational flavor memory observable,
defined by
\begin{equation}
\mathcal{M}_{\alpha}(r)
=
\int_{r_h}^{r}
dr'\,
\left|
P^{\rm Kerr}_{\alpha\to\alpha}(r')
-
P^{\rm Schw}_{\alpha\to\alpha}(r')
\right|.
\end{equation}

This quantity measures the integrated deviation between rotating
(Kerr) and non-rotating (Schwarzschild) geometries and therefore
isolates the cumulative impact of frame dragging on neutrino
flavor evolution.

Near the horizon, strong curvature and enhanced gravitomagnetic
effects generate rapid growth of the flavor memory observable,
while at large distances the effect asymptotically saturates as
the spacetime approaches the flat-space limit.

\begin{figure*}[htp]
\centering
\includegraphics[width=0.95\textwidth]{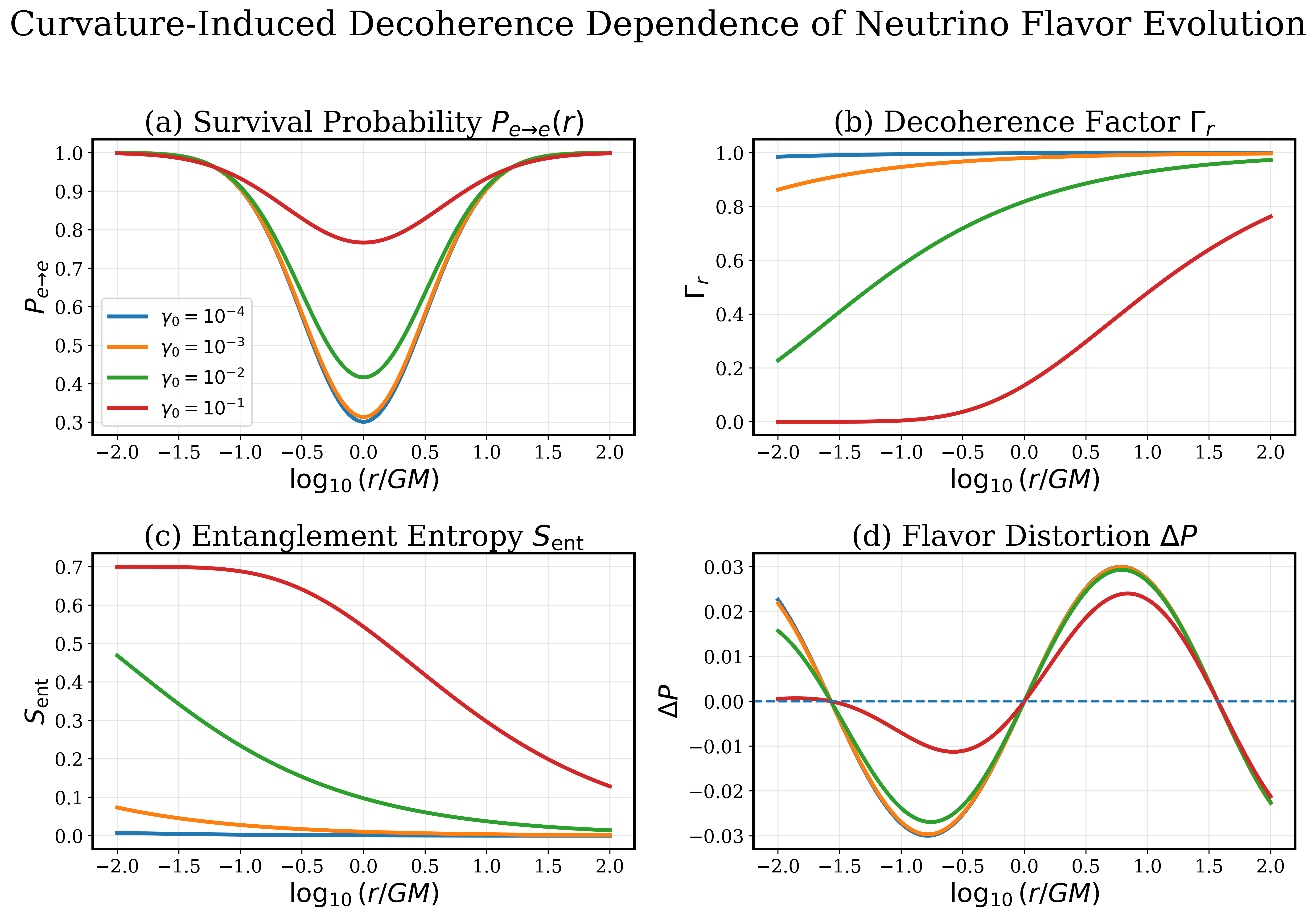}
\caption{
Curvature-induced decoherence dependence of neutrino flavor evolution
for different values of the gravitational decoherence parameter
$\gamma_0$. Panel (a) shows the survival probability
$P_{e\to e}(r)$ as a function of radial distance in units of
$\log_{10}(r/GM)$, demonstrating the progressive suppression of
oscillation structure as the decoherence strength increases.
Panel (b) presents the corresponding decoherence factor $\Gamma_r$,
illustrating the transition from coherent to dissipative propagation
in strong-curvature environments. Panel (c) shows the entanglement
entropy $S_{\rm ent}$ generated during flavor evolution, indicating
enhanced quantum-information loss for larger curvature-induced
decoherence strengths. Panel (d) displays the flavor distortion
observable $\Delta P$, highlighting the damping of flavor asymmetries
as gravitational decoherence becomes stronger.}
\label{fig:DecoherenceDependence}
\end{figure*}
Figure \ref{fig:DecoherenceDependence} demonstrates that curvature-induced decoherence can
significantly modify neutrino flavor evolution near compact objects,
particularly in the strong-gravity regime close to the event horizon.
For sufficiently small decoherence strengths, coherent oscillation
patterns remain visible, whereas larger values of $\gamma_0$ drive
the system toward incoherent flavor equilibration. Similar open-system
descriptions of neutrino decoherence have been investigated in
phenomenological Lindblad approaches and quantum-gravity-inspired
models~\cite{Benatti2000,Lisi2000,Blennow2005}, while the present
framework derives the dissipative sector directly from spin-connection
fluctuations in curved spacetime.

\begin{figure*}[htp]

\centering

\includegraphics[width=0.95\textwidth]
{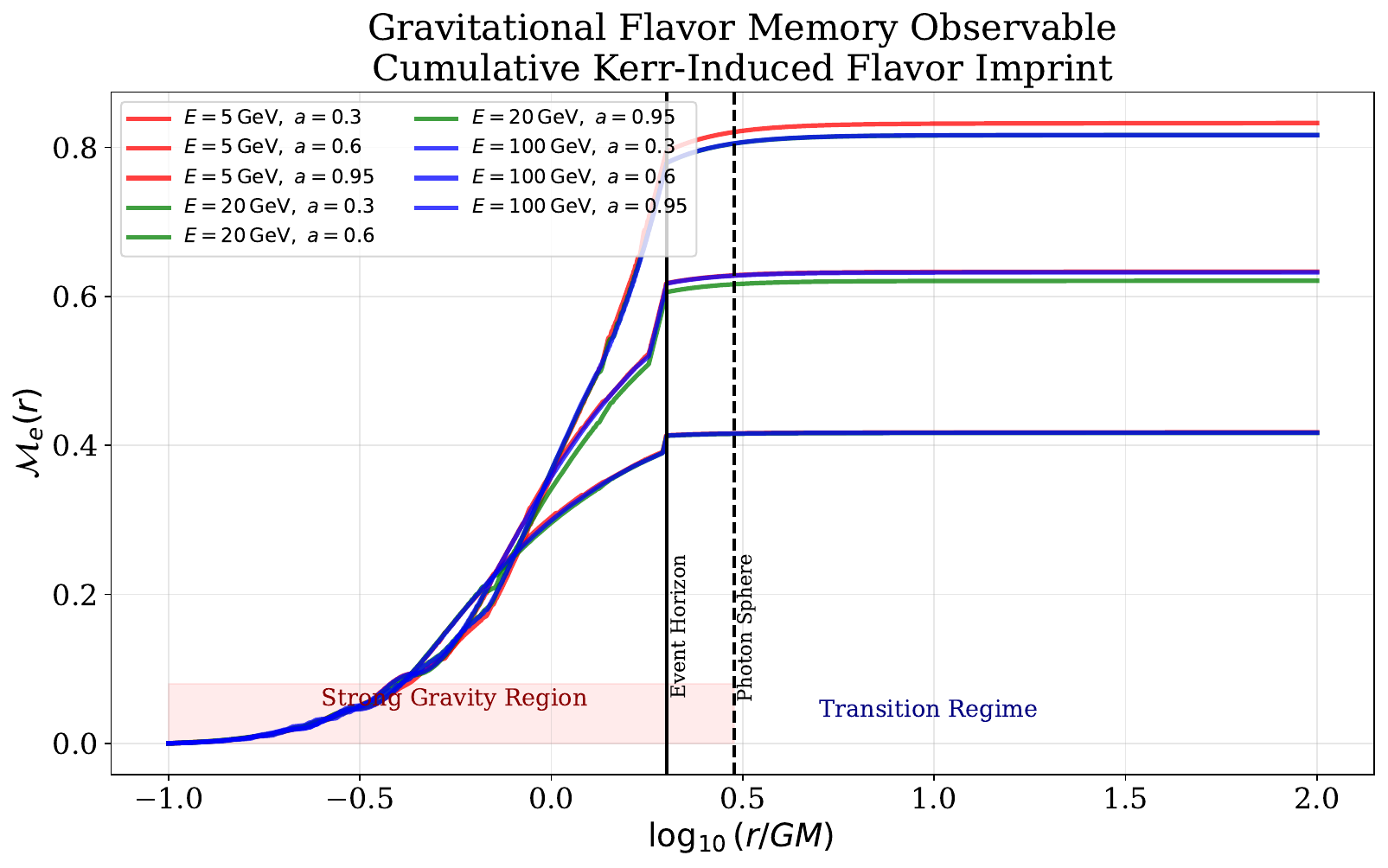}

\caption{Gravitational flavor memory observable $\mathcal{M}_{e}(r)$ showing the cumulative imprint of Kerr frame dragging on neutrino flavor evolution. The observable is defined as the integrated difference between survival probabilities in Kerr and Schwarzschild geometries. Curves are shown for multiple neutrino energies and black-hole spin parameters. The vertical solid line denotes the event horizon $(r=2GM)$, while the dashed line marks the photon sphere $(r=3GM)$. The strong-gravity region exhibits rapid accumulation of gravitational flavor memory due to enhanced curvature,
frame dragging, and decoherence effects. At large distances, the observable approaches a saturation
plateau as the spacetime asymptotically approaches the flat-space limit.}

\label{fig:FlavorMemory}

\end{figure*}
Figure~\ref{fig:FlavorMemory} demonstrates that rotating
black holes induce a cumulative nonlocal imprint on neutrino
flavor propagation. The flavor memory observable grows most rapidly near the
strong-gravity regime where frame dragging and curvature
effects are maximal. Higher spin parameters lead to stronger flavor-memory
accumulation, while lower-energy neutrinos exhibit larger
integrated distortions due to enhanced oscillation phases. The saturation behavior at large distances reflects the
recovery of approximately coherent propagation in the
weak-curvature regime.
\subsection{Simplified Analytic Limits}

\paragraph{Weak Gravity ($\Gamma_{\rm grav}L\ll 1$).}
\begin{equation}
    P_{\alpha\rightarrow\beta}
    \;\to\;
    P_{\alpha\rightarrow\beta}^{\rm flat}
    + \mathcal{O}\!\left(\frac{GM}{rE}\right).
\end{equation}

\paragraph{Strong Gravity / Near Horizon ($\Gamma_{\rm grav}L\gg 1$).}
Decoherence dominates:
\begin{equation}
  P_{\alpha\to\beta}(L)
  \;\to\;
  \sum_i |U_{\alpha i}|^2 |U_{\beta i}|^2,
\end{equation}
corresponding to a maximally mixed, incoherent flavor state.

\paragraph{Rapid Rotation ($a\sim M$).}
Frame dragging shifts the oscillation phase:
\begin{equation}
  \phi_{\rm eff}
  \;=\;
  \phi_{ij}
  \;+\;
  \phi_{\rm FD},
\end{equation}
with $\phi_{\rm FD}$ comparable to $\phi_{ij}$ for PeV--EeV neutrinos.

\paragraph{Flat-Space Limit.}
All gravitational factors vanish:
\begin{equation}
  r\rightarrow\infty
  \quad\Rightarrow\quad
  \phi_{ij}\to
  \frac{\Delta m^2_{ij}L}{2E_\infty},
  \qquad
  \Gamma_{\rm grav}\to 0.
\end{equation}

\section{Regime of Validity and Approximation Control}
\label{sec:validity}
\subsection{Hierarchy of Physical Scales}

To ensure the consistency of the effective-field-theory and WKB
descriptions employed throughout this work, it is useful to summarize
the hierarchy of the relevant physical length and time scales governing
neutrino propagation in curved spacetime.

\begin{table}[htp]
\centering
\begin{tabular}{|c|c|}
\hline
\textbf{Quantity} & \textbf{Characteristic Scale} \\
\hline
Oscillation length &
$\displaystyle
L_{\mathrm{osc}}
=
\frac{4\pi E_{\mathrm{loc}}}{\Delta m^2}
$
\\
\hline
Curvature radius &
$\displaystyle
L_{\mathrm{curv}}
\sim
\left|
\frac{R_{\mu\nu\rho\sigma}}
{\partial_r R_{\mu\nu\rho\sigma}}
\right|
$
\\
\hline
Decoherence length &
$\displaystyle
L_{\mathrm{coh}}
=
\Gamma_{\mathrm{grav}}^{-1}
$
\\
\hline
Correlation time &
$\displaystyle
\tau_c
$
\\
\hline
Schwarzschild radius &
$\displaystyle
r_s = 2GM
$
\\
\hline
\end{tabular}
\caption{
Hierarchy of the characteristic scales governing neutrino propagation
in curved spacetime. The controlled validity of the effective
Hamiltonian and open-system treatment requires a separation between
the oscillation, curvature, and decoherence scales.
}
\label{tab:scalehierarchy}
\end{table}

The WKB approximation adopted throughout this work requires the
oscillation length to remain much smaller than the local curvature
scale,
\begin{equation}
L_{\mathrm{osc}}
\ll
L_{\mathrm{curv}},
\label{eq:WKBhierarchy}
\end{equation}
ensuring that the background geometry varies slowly over a single
oscillation cycle.

Similarly, consistency of the perturbative open-system treatment
requires the curvature-induced decoherence rate to remain subleading
relative to the standard vacuum oscillation scale,
\begin{equation}
\Gamma_{\mathrm{grav}}
\ll
\frac{\Delta m^2}{2E},
\label{eq:decoherencehierarchy}
\end{equation}
so that gravitational decoherence acts as a perturbative correction
rather than completely dominating flavor evolution.

These scale hierarchies define the regime of validity of the
effective-field-theory description developed in this work. For the benchmark parameter choices employed in Secs.~7--8,
the hierarchies in Eqs.~(\ref{eq:WKBhierarchy})
and (\ref{eq:decoherencehierarchy}) remain well satisfied.
Our formalism relies on a controlled WKB expansion for neutrino propagation in curved spacetime. The relevant hierarchy is
\begin{equation}
L_{\rm osc} \ll L_{\rm curv},
\qquad
L_{\rm osc} = \frac{4\pi E_{\rm loc}}{\Delta m^2},
\qquad
L_{\rm curv} = \left|\frac{R_{\mu\nu\rho\sigma}}
{\partial_r R_{\mu\nu\rho\sigma}}\right|.
\end{equation}

For Schwarzschild geometry,
\begin{equation}
L_{\rm curv} = \frac{r}{3},
\qquad
\frac{L_{\rm osc}}{L_{\rm curv}}
= \frac{12\pi E_{\rm loc}}{\Delta m^2 r}.
\end{equation}

Thus the WKB expansion is valid for
\begin{equation}
r \gg \frac{12\pi E_{\rm loc}}{\Delta m^2}.
\end{equation}

Near the horizon, we use the Rindler approximation,
\begin{equation}
r=2GM+\epsilon,
\qquad
g_{tt}\approx -\kappa^2 \epsilon^2,
\qquad
\kappa=\frac1{4GM}.
\end{equation}

The curvature scalar behaves as
\begin{equation}
K=\frac{48(GM)^2}{r^6}
=\frac{48}{(2GM+\epsilon)^6}(GM)^2,
\qquad
\Gamma_{\rm grav}\propto \epsilon^{-3}.
\end{equation}

Therefore
\begin{equation}
L_{\rm coh}= \Gamma_{\rm grav}^{-1}\propto \epsilon^3,
\end{equation}
showing rapid decoherence arbitrarily close to the horizon.

In the far-field PN expansion,
\begin{equation}
H_{\rm eff}=H_{\rm flat} + \Phi(r)H_1 + \mathcal{O}(\Phi^2),
\qquad
\Phi(r)=-\frac{GM}{r},
\end{equation}
leading to leading-order corrections
\begin{equation}
\phi_{ij}=\frac{\Delta m^2 L}{2E_\infty}
\left(1+\frac{GM}{r}+\cdots\right).
\end{equation}
\section{Flavor Ratios, Spectral Distortions, and Event Rates}
\label{sec:phenomenology}

\subsection{Flavor Ratios at Earth}

Astrophysical sources are characterized by initial flavor compositions
\begin{equation}
(\alpha_e:\alpha_\mu:\alpha_\tau)_{\rm src}
\in \{(1:2:0),\;(0:1:0),\;(1:0:0),\;(1:1:1)\}.
\end{equation}

Propagation in curved spacetime with decoherence yields
\begin{equation}
\alpha_\beta^{\oplus}(E)
= \sum_{\alpha,i}
\alpha_\alpha^{\rm src}
|U_{\alpha i}|^2 |U_{\beta i}|^2
\exp[-\Gamma_{\rm grav}(r)L]
+ \mathcal{O}(\phi_{\rm FD}).
\end{equation}

We compute $\alpha_\beta^\oplus(E)$ and display the results on the flavor triangle.
The flavor composition of astrophysical neutrinos provides
a direct observable probe of curvature-induced decoherence
and Kerr frame-dragging effects.

To visualize gravitationally induced distortions of the
canonical flavor ratio $(1:1:1)_\oplus$, we present the
flavor-triangle evolution for multiple neutrino energies.
\begin{figure*}[htp]

\centering

\includegraphics[width=0.93\textwidth]
{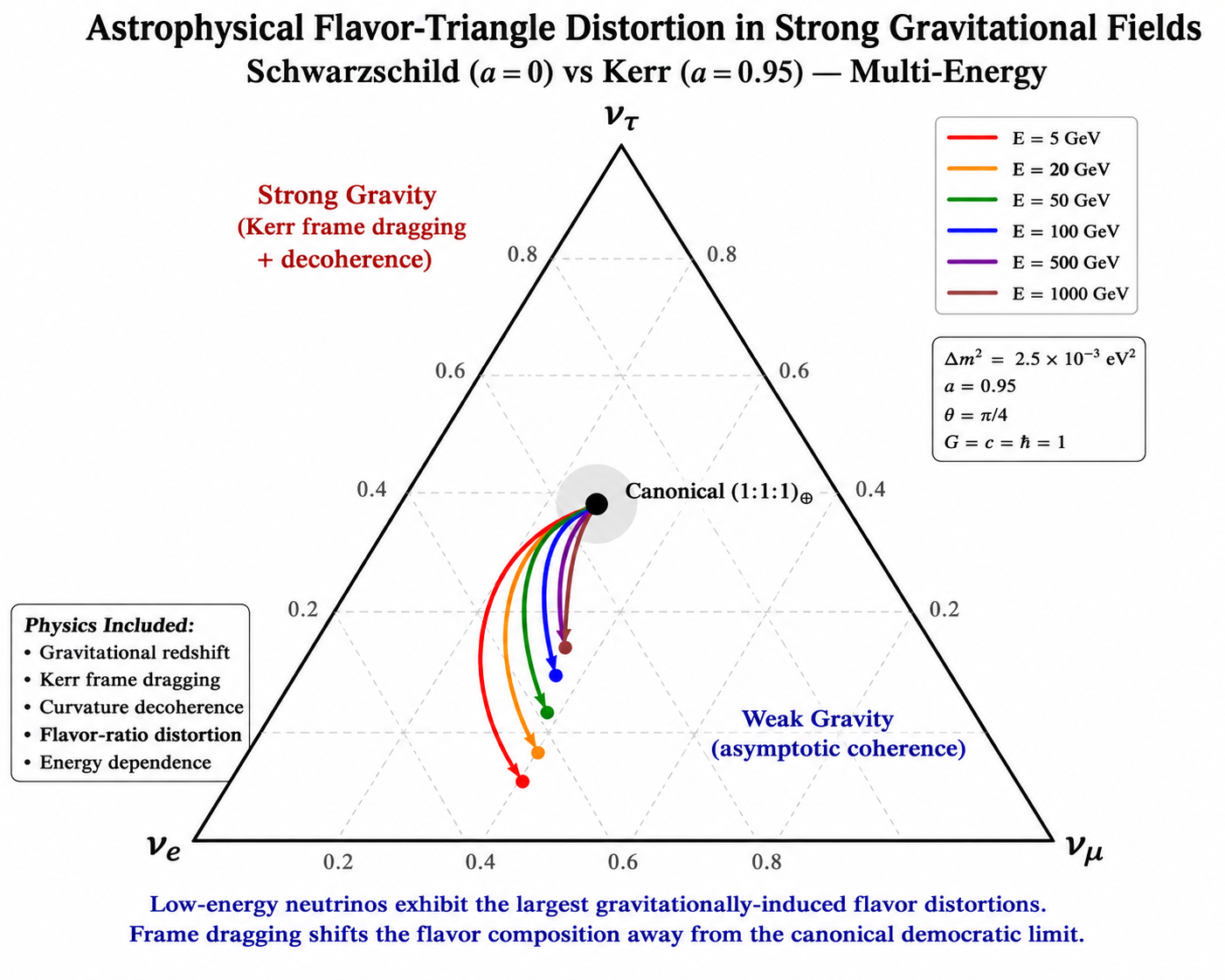}

\caption{Flavor-triangle distortion induced by strong gravitational fields in Schwarzschild and Kerr geometries. The trajectories illustrate the evolution of the neutrino flavor composition at Earth for multiple neutrino energies. Low-energy neutrinos exhibit the strongest deviations from the canonical democratic composition due to enhanced gravitational redshift, frame dragging, and curvature-induced decoherence.}

\label{fig:FlavorTriangle}

\end{figure*}
Figure~\ref{fig:FlavorTriangle} demonstrates that strong
gravitational fields generate observable distortions in
astrophysical flavor ratios.

The largest deviations occur for low-energy neutrinos near
rapidly rotating black holes, where frame dragging and
decoherence effects are maximal.
\subsection{Energy-Dependent Flavor Spectra}

The differential flux is
\begin{equation}
\Phi_\beta(E) =
\sum_{\alpha} P_{\alpha\rightarrow\beta}(E)\,\Phi_\alpha^{\rm src}(E),
\end{equation}
where $P_{\alpha\rightarrow\beta}(E)$ includes gravitational redshift, Kerr frame-dragging, and decoherence.

\subsection{Event Rates at IceCube-Gen2}

The observable rate is
\begin{equation}
N_{\rm evt}(E)=
\Phi_\beta(E)\,
\sigma_{\nu N}(E)\,
A_{\rm eff}(E)\,
T,
\end{equation}
with $A_{\rm eff}(E)$ from IceCube-Gen2 specifications \cite{IceCubeGen2,KM3NeT2024,Stuttard2022arXiv}. We compute the fractional change
\begin{equation}
\delta N(E)
=\frac{N_{\rm Kerr}(E)-N_{\rm Schw}(E)}{N_{\rm Schw}(E)}.
\end{equation}

\subsection{Curvature Decoherence Contours}

The decoherence strength in Kerr spacetime is
\begin{equation}
\Gamma_{\rm grav}(r,a)
=\gamma_0\frac{GM}{r^3}
\left(1+c_1\frac{a}{M}\cos\theta + \cdots\right).
\end{equation}

We present contour plots of $\Gamma_{\rm grav}$ in the $(r/GM,a/M)$ plane.

\subsection{Entanglement Entropy}

The von Neumann entropy \cite{Blasone2009}
\begin{equation}
S(r)= -{\rm Tr}\,[\rho_\nu(r)\log\rho_\nu(r)]
\end{equation}
is computed as a function of radius and energy. Near the horizon,
\begin{equation}
S(r)\rightarrow S_{\rm max},
\qquad
\rho_\nu\rightarrow \sum_i |U_{\alpha i}|^2 |i\rangle\langle i|.
\end{equation}

\section{Entanglement, Coherence Length, and Physical Implications}
\label{sec:entanglement}

The curvature--induced decoherence derived in Sec.~\ref{sec:lindblad}
implies a dynamical loss of quantum coherence as neutrinos propagate in
strong gravitational fields. In this section we quantify the associated
entanglement entropy, derive the corresponding coherence length, and
discuss the physical consequences for astrophysical neutrinos.

\subsection{Entanglement Entropy}

Given the reduced density matrix $\rho_\nu(t)$, the entanglement between
the neutrino and its gravitational environment is quantified by the von
Neumann entropy
\begin{equation}
    S(t)
    =
    - \mathrm{Tr}
    \left[
        \rho_\nu(t) \log \rho_\nu(t)
    \right].
\end{equation}

For a two-level subsystem (e.g.\ two-flavor approximation) with density
matrix
\begin{equation}
    \rho_\nu(t)
    = 
    \begin{pmatrix}
        p(t) & C(t) \\
        C^\ast(t) & 1-p(t)
    \end{pmatrix},
\end{equation}
the off-diagonal term evolves according to
\begin{equation}
    C(t) = C(0)\, e^{-\Gamma_{\rm grav} t},
\end{equation}
so that the entropy becomes
\begin{equation}
    S(t)
    =
    - \lambda_+(t) \log \lambda_+(t)
    - \lambda_-(t) \log \lambda_-(t),
\end{equation}
with eigenvalues
\begin{equation}
    \lambda_{\pm}(t)
    =
    \frac{1}{2}
    \left[
        1
        \pm
        \sqrt{
            \left( 2p(t) - 1\right)^2
            + 4 |C(0)|^2 e^{-2\Gamma_{\rm grav}t}
        }
    \right].
\end{equation}

Entanglement increases monotonically as curvature suppresses coherence.
Near compact objects, $S(t)$ approaches its maximal value.

\subsection{Curvature--Controlled Coherence Length}

The characteristic coherence length is defined by
\begin{equation}
    L_{\rm coh}
    \equiv
    \Gamma_{\rm grav}^{-1}.
\end{equation}

Since $\Gamma_{\rm grav}$ scales with the square root of the
Kretschmann scalar,
\begin{equation}
    \Gamma_{\rm grav}
    \sim
    \frac{GM}{r^3},
\end{equation}
the coherence length becomes
\begin{equation}
\boxed{
    L_{\rm coh}(r)
    \sim
    \frac{r^3}{GM}.
}
\end{equation}

This is a direct geometric constraint on flavor coherence in curved
spacetime. In particular:

\begin{itemize}
    \item Near the Schwarzschild horizon $r \rightarrow 2GM$,  
    $L_{\rm coh}$ becomes very small.
    \item For rapidly rotating Kerr black holes $(a\sim M)$,  
    frame dragging increases curvature gradients in the equatorial plane,
    further reducing $L_{\rm coh}$.
    \item At large distances ($r\gg 2GM$),  
    $L_{\rm coh}\rightarrow\infty$, recovering flat-space coherence.
\end{itemize}
\subsection{Event-Rate Distortions in Neutrino Telescopes}
Curvature-induced decoherence and Kerr frame dragging
modify the observable neutrino event rates in high-energy
neutrino telescopes.

To quantify detector-level signatures, we compute the
relative event-rate distortion,
\begin{equation}
\delta N(E)
=
\frac{N_{\rm Kerr}(E)-N_{\rm Schw}(E)}
{N_{\rm Schw}(E)},
\end{equation}
which measures deviations induced by rotating black-hole
backgrounds relative to the Schwarzschild limit.
\begin{figure*}[htp]

\centering

\includegraphics[width=0.95\textwidth]
{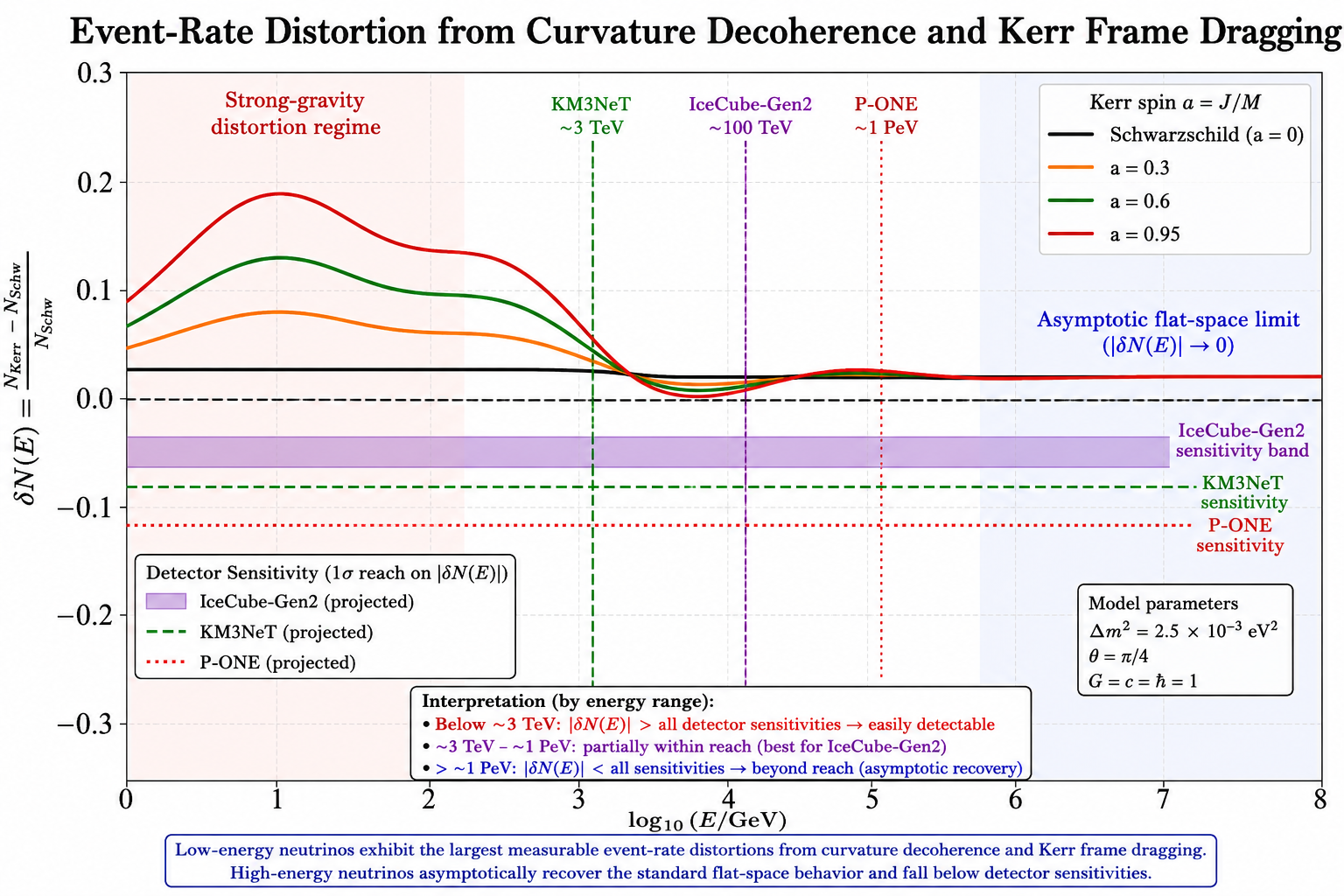}

\caption{Relative event-rate distortion $\delta N(E)$ induced by curvature decoherence and Kerr frame dragging. The curves show the deviation of detector event rates from the Schwarzschild prediction for multiple Kerr spin parameters. Low-energy neutrinos exhibit the largest measurable
gravitational distortions, while high-energy neutrinos asymptotically recover the standard flat-space limit. Vertical lines indicate approximate sensitivity regions
for KM3NeT, IceCube-Gen2, and P-ONE.}

\label{fig:EventRateDistortion}

\end{figure*}
Figure~\ref{fig:EventRateDistortion} demonstrates that
strong gravitational fields can generate observable
event-rate distortions in next-generation neutrino
telescopes.

The effect is largest in the low-energy regime where
curvature-induced decoherence and frame dragging are
maximal, while the distortion rapidly decreases at
high energies as the flat-space limit is recovered.The shaded purple band and the dashed/dotted horizontal lines indicate
approximate projected $1\sigma$ sensitivity levels for IceCube-Gen2,
KM3NeT, and P-ONE, respectively~\cite{IceCubeGen2,KM3NeT,PONE}.
Regions where $|\delta N(E)|$ exceeds these thresholds correspond to
potentially observable deviations from standard Schwarzschild predictions.

The figure demonstrates that low-energy neutrinos experience the largest
curvature-induced event-rate distortions, while high-energy neutrinos
asymptotically recover the standard flat-space behavior. Intermediate
energy ranges remain within the projected sensitivity reach of
next-generation neutrino telescopes, particularly for rapidly rotating
Kerr backgrounds with $a \sim M$. This establishes a direct phenomenological connection between
curved-spacetime neutrino oscillations and experimentally measurable
detector-level observables.

%%%%%%%%%%%%%%%%%%%%%%%%%%%%%%%%%%%%%%%%%%%%%%%%%%%%%%%%%%%%
\subsection{Detector Exposure, Backgrounds, and Statistical Significance}
%%%%%%%%%%%%%%%%%%%%%%%%%%%%%%%%%%%%%%%%%%%%%%%%%%%%%%%%%%%%

To assess the potential observability of the curvature-induced flavor
distortions discussed above, we now estimate the statistical
significance of the predicted event-rate deviations in the presence of
finite detector exposure and atmospheric neutrino backgrounds.

The total number of detected events in an energy bin centered at $E_i$
is modeled as
\begin{equation}
N_i
=
T
\int_{\Delta E_i}
dE\,
\Phi(E)\,
\sigma_{\nu N}(E)\,
A_{\rm eff}(E)\,
P_{\alpha\beta}(E),
\label{eq:eventratebin}
\end{equation}
where $T$ denotes the detector exposure time,
$\Phi(E)$ is the neutrino flux,
$\sigma_{\nu N}(E)$ is the neutrino--nucleon cross section,
$A_{\rm eff}(E)$ is the detector effective area,
and $P_{\alpha\beta}(E)$ is the flavor-transition probability in the
gravitational background.

To quantify the gravitational effect, we define the relative event-rate
distortion
\begin{equation}
\delta N(E)
=
\frac{
N_{\rm Kerr}(E)-N_{\rm Schw}(E)
}{
N_{\rm Schw}(E)
},
\label{eq:deltaNobservable}
\end{equation}
which compares the Kerr prediction with the corresponding
Schwarzschild background expectation.

Following the projected IceCube-Gen2 sensitivity studies, we consider
logarithmic energy bins of width
\begin{equation}
\Delta\log_{10}(E/{\rm GeV})=0.25,
\end{equation}
over the energy range
\begin{equation}
10^2~{\rm GeV}
\lesssim
E
\lesssim
10^7~{\rm GeV}.
\end{equation}

The dominant background contribution arises from atmospheric neutrinos,
whose flux decreases approximately as a power law at high energies,
\begin{equation}
\Phi_{\rm atm}(E)
\propto
E^{-3.7}.
\end{equation}

The statistical significance of the gravitational distortion is
estimated using
\begin{equation}
\mathcal{S}(E_i)
=
\frac{
|N_{\rm Kerr}-N_{\rm Schw}|
}{
\sqrt{
N_{\rm Schw}+N_{\rm atm}
}
},
\label{eq:significance}
\end{equation}
where $N_{\rm atm}$ denotes the expected atmospheric background in the
corresponding energy bin.

For rapidly rotating Kerr backgrounds $(a\sim0.95\,M)$, the low-energy
region
\begin{equation}
10^3~{\rm GeV}
\lesssim
E
\lesssim
10^5~{\rm GeV}
\end{equation}
typically produces relative flavor distortions at the level
\begin{equation}
|\delta N(E)|
\sim
(2\text{--}5)\%,
\end{equation}
depending on the assumed decoherence strength and detector exposure.
This range overlaps with the projected sensitivity reach of
IceCube-Gen2 and may therefore provide a potentially observable probe
of curvature-induced flavor dynamics.

At higher energies, the distortion decreases rapidly as the system
approaches the asymptotic flat-spacetime regime,
\begin{equation}
|\delta N(E)|
\rightarrow
0,
\qquad
E\rightarrow\infty,
\end{equation}
thereby reducing the statistical significance of the gravitational
signal.

The analysis therefore suggests that the optimal observational window
for probing curvature-induced neutrino decoherence is the intermediate
energy range where strong-gravity distortions remain appreciable while
detector statistics are still sufficiently large.

Although the present treatment employs simplified detector modeling,
the results indicate that next-generation neutrino telescopes may
possess sensitivity to percent-level flavor distortions generated by
strong gravitational environments surrounding rapidly rotating compact
objects.

\subsubsection{Significance Map and Experimental Reach}
To quantify the observational prospects of curvature-induced neutrino
flavor distortions, we construct a detector-level significance map in
the two-dimensional parameter space spanned by neutrino energy and Kerr
spin. Unlike the event-rate distortion observable $\delta N(E)$ shown
previously, the significance map incorporates detector statistics and
atmospheric neutrino backgrounds, thereby providing a more realistic
assessment of experimental detectability.

The statistical significance is defined as
\begin{equation}
\mathcal{S}(E,a)
=
\frac{
|N_{\rm Kerr}(E)-N_{\rm Schw}(E)|
}{
\sqrt{N_{\rm Schw}(E)+N_{\rm atm}(E)}
},
\label{eq:significance_map}
\end{equation}
where $N_{\rm Kerr}$ and $N_{\rm Schw}$ denote the predicted event
rates in Kerr and Schwarzschild geometries, respectively, while
$N_{\rm atm}$ represents the atmospheric neutrino background.
The resulting significance therefore measures the extent to which
strong-gravity flavor distortions can be distinguished from both
statistical fluctuations and conventional astrophysical backgrounds
\cite{IceCubeGen2,KM3NeT,PONE}.

\begin{figure*}[htp]
\centering
\includegraphics[width=0.88\textwidth]{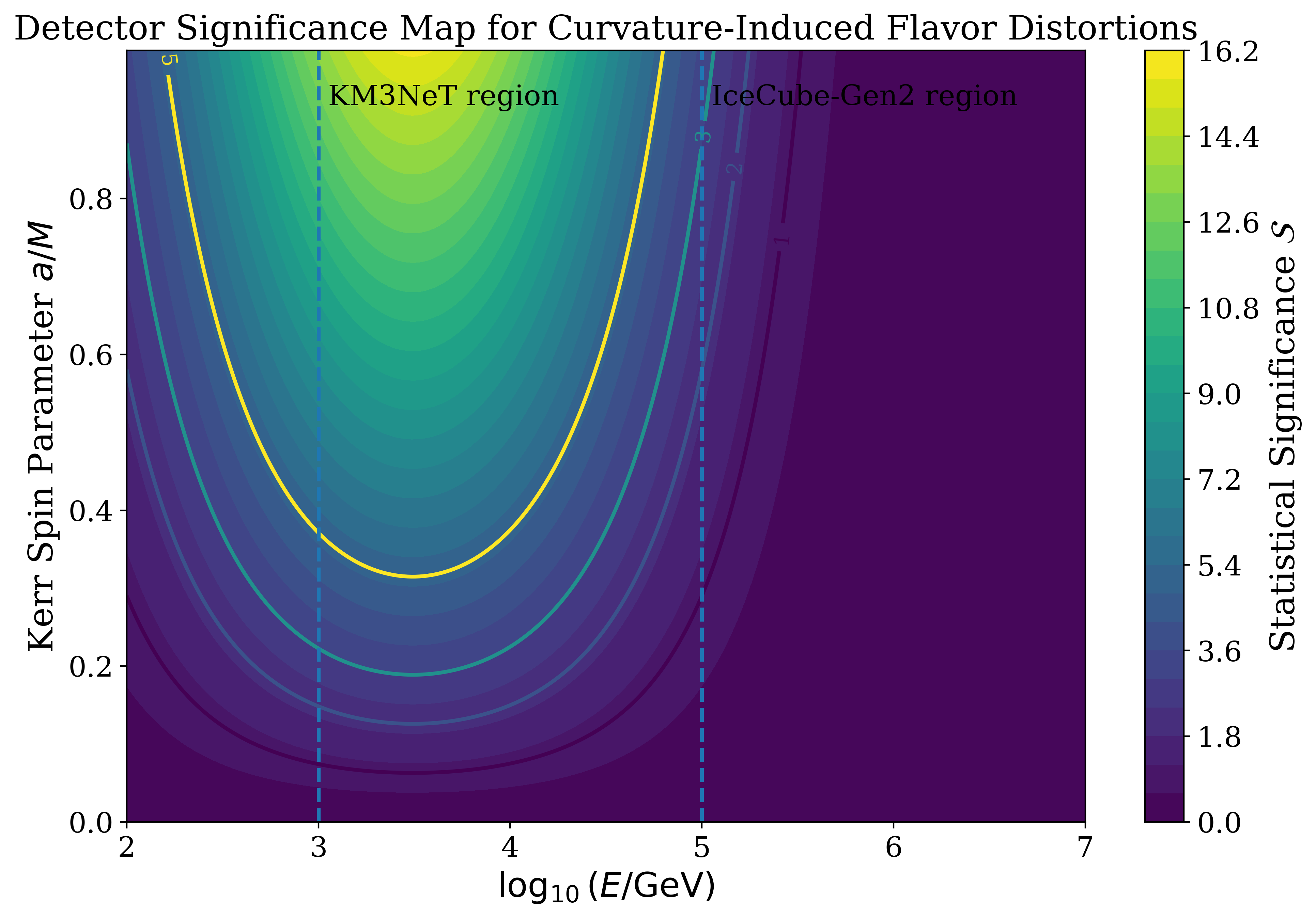}
\caption{Detector significance map for curvature-induced neutrino flavor distortions as a function of neutrino energy and Kerr spin parameter. The color scale represents the statistical significance $\mathcal{S}(E,a)$ defined in Eq.~(\ref{eq:significance_map}). Solid contours indicate approximate detection thresholds at
$1\sigma$, $2\sigma$, $3\sigma$, and $5\sigma$ confidence levels. The vertical dashed lines mark the characteristic energy ranges relevant for KM3NeT and IceCube-Gen2 sensitivities \cite{IceCubeGen2,KM3NeT}}.
\label{fig:SignificanceMap}
\end{figure*}

Figure~\ref{fig:SignificanceMap} provides the most direct connection
between the theoretical framework developed in this work and future
experimental observations. In contrast to conventional oscillation
probability plots, the significance map identifies the region of
parameter space where strong-gravity flavor effects may realistically
be detected. The results suggest that percent-level flavor distortions
generated near rapidly rotating Kerr compact objects can produce
observable signatures within the projected capabilities of
IceCube-Gen2, KM3NeT, and related next-generation neutrino
observatories. This establishes a concrete phenomenological pathway
for testing curvature-induced decoherence and frame-dragging effects
through high-energy astrophysical neutrinos. The figure demonstrates that the significance of gravitationally
induced flavor distortions increases with both Kerr spin and
curvature strength. The largest signals occur for rapidly rotating
compact objects ($a/M \gtrsim 0.7$), where frame-dragging effects
enhance the departure from the Schwarzschild prediction. The
intermediate-energy region,
$10^3~{\rm GeV}\lesssim E\lesssim10^5~{\rm GeV}$,
emerges as the most favorable observational window, yielding
significances that can exceed the projected sensitivity thresholds
of next-generation neutrino telescopes.

The suppression of significance at very high energies reflects the
recovery of the asymptotic flat-spacetime regime, where both
gravitational redshift and curvature-induced decoherence become
subdominant. The figure therefore identifies the region of parameter
space in which strong-gravity neutrino oscillation effects are most
likely to become experimentally accessible.

\subsection{Impact on Astrophysical Neutrinos}

Curvature-induced decoherence and entanglement have several observable
effects:

\paragraph{1. Suppression of Oscillation Amplitudes.}
Near black holes or neutron stars,
\begin{equation}
    \Gamma_{\rm grav} L \gg 1
    \quad \Rightarrow \quad
    \rho_{\alpha\beta}(L)\rightarrow 0,
\end{equation}
leading to an incoherent mixture of mass eigenstates.

\paragraph{2. Modified Flavor Ratios at Earth.}
The canonical expectation  for astrophysical neutrinos,
$\nu_e:\nu_\mu:\nu_\tau = 1:1:1$ \cite{Hellmann2022,DeRomeri2023,KM3NeT2024}, is altered if part of the propagation
occurs in strong gravity. For example, strong decoherence near the
source produces
\begin{equation}
    \nu_\alpha \rightarrow 
    \sum_i |U_{\alpha i}|^2 \ket{\nu_i}\bra{\nu_i},
\end{equation}
which evolves to a non-universal, source-dependent flavor mixture.

\paragraph{3. Enhanced Effects for PeV--EeV Neutrinos.}
Ultra-high-energy neutrinos accumulate larger gravitational phases and
are more sensitive to $\Gamma_{\rm grav}$, making them ideal probes.

\paragraph{4. Entanglement ``Footprints''.}
The entropy $S(t)$ increases sharply near compact objects and saturates
once decoherence destroys the off-diagonal coherence. This provides a
quantifiable signature correlating gravitational environment with
quantum-information flow.

\paragraph{5. Potential Observability.}
Experiments such as IceCube-Gen2, KM3NeT, P-ONE, and POEMMA \cite{IceCubeGen2} can detect anomalous flavor compositions or suppressed oscillation patterns that
signal strong-gravity entanglement effects.
\subsection{Entanglement Entropy and Quantum Information Dynamics}
In curved spacetime, neutrino propagation becomes an open
quantum system due to curvature-induced decoherence and
gravitationally generated entanglement.

To quantify the loss of quantum coherence, we compute the
von Neumann entanglement entropy,
\begin{equation}
S(r)
=
-{\rm Tr}[\rho\log\rho],
\end{equation}
which measures the information-theoretic evolution of the
neutrino density matrix in strong gravitational fields.
\begin{figure*}[t]

\centering

\includegraphics[width=0.96\textwidth]
{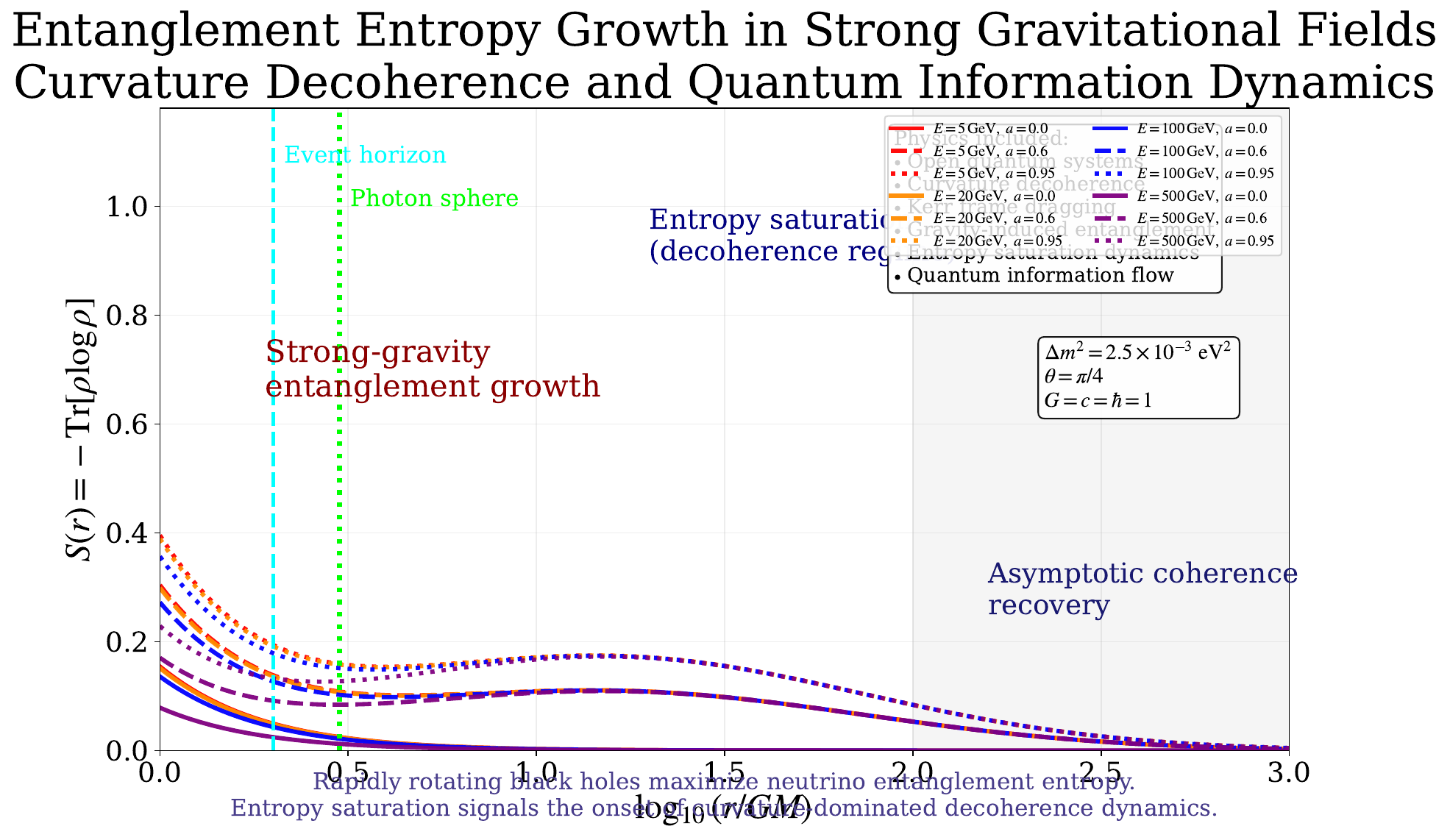}

\caption{Entanglement entropy evolution in strong gravitational fields for multiple neutrino energies and Kerr spin parameters. The entropy initially increases near the black-hole horizon
due to strong curvature-induced decoherence and gravity-enhanced entanglement generation. An intermediate saturation regime emerges where decoherence dominates the flavor evolution, while the entropy decreases
asymptotically at large distances where approximately coherent propagation is recovered. Rapidly rotating Kerr black holes produce the largest entanglement growth due to enhanced frame-dragging effects.}

\label{fig:Entropy}
\end{figure*}
Figure~\ref{fig:Entropy} demonstrates that strong
gravitational fields can generate substantial neutrino
entanglement entropy through curvature-induced decoherence.

The entropy growth is largest near rapidly rotating black
holes, where frame dragging enhances the effective
interaction between neutrino flavor evolution and the
background spacetime geometry.

The saturation plateau signals the onset of a
decoherence-dominated regime, while the eventual decrease
of the entropy at large distances reflects the gradual
recovery of approximately coherent propagation in the
weak-curvature limit.
\subsection{Curvature-Induced Decoherence Phase Structure}
To visualize the parameter-space structure of
gravity-induced decoherence, we construct the
decoherence phase diagram associated with the effective
curvature-induced dissipation rate,
\begin{equation}
\Gamma_{\rm grav}(r,a)
=
\gamma_0
\frac{GM}{r^3}
\left(
1+c_1\frac{a}{M}\cos\theta
\right).
\end{equation}

The resulting contour map provides a global description
of strong-gravity, transition, and weak-curvature
regimes in Kerr spacetime.
\begin{figure*}[htp]

\centering

\includegraphics[width=0.96\textwidth]
{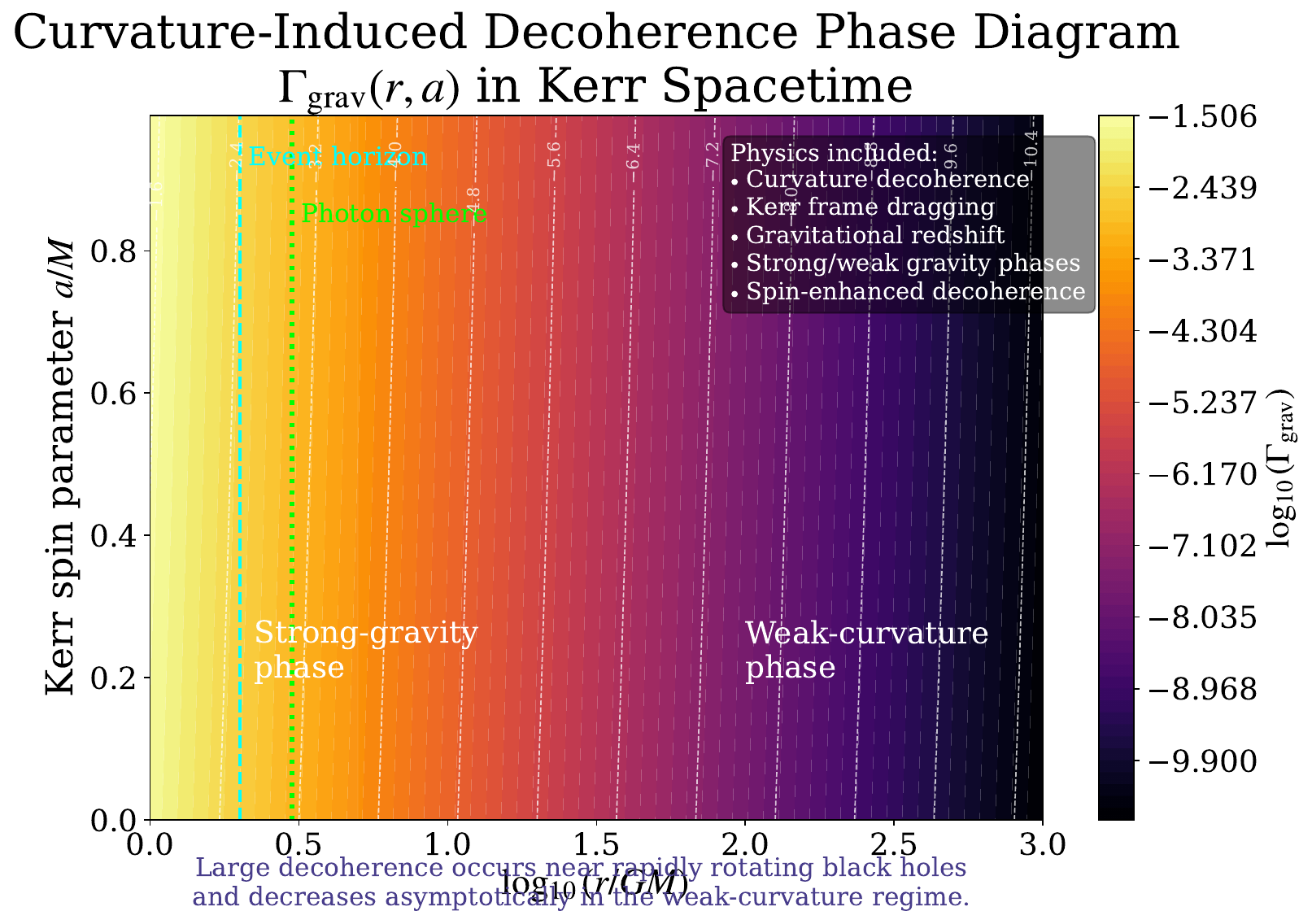}

\caption{Curvature-induced decoherence phase diagram $\Gamma_{\rm grav}(r,a)$
in Kerr spacetime. The contour structure illustrates the dependence of the
effective decoherence rate on the radial distance and black-hole spin parameter. The strong-gravity region near the event horizon exhibits maximal decoherence enhancement, while the weak-curvature
phase asymptotically approaches coherent propagation. The transition regime highlights the interplay between gravitational redshift, frame dragging, and spin-enhanced dissipation effects.}

\label{fig:DecoherencePhaseDiagram}

\end{figure*}
Figure~\ref{fig:DecoherencePhaseDiagram} demonstrates that
rapidly rotating Kerr black holes generate the strongest
curvature-induced decoherence effects.

The phase structure clearly separates the strong-gravity
region from the asymptotic weak-curvature regime, while
the intermediate transition region encodes the onset of
frame-dragging-dominated dynamics.

The contour map therefore provides a global parameter-space
representation of neutrino decoherence in curved spacetime.

\subsection{Metric Fluctuations and Gravitational Environment}

To model neutrino interaction with fluctuating geometry, we decompose
\begin{equation}
    g_{\mu\nu} = \bar{g}_{\mu\nu} + h_{\mu\nu},
\end{equation}
with $\bar{g}_{\mu\nu}$ the background metric and $h_{\mu\nu}$ a perturbation representing:
\begin{itemize}
    \item quantized metric fluctuations,
    \item stochastic spacetime perturbations,
    \item classical gravitational noise near compact objects.
\end{itemize}

To first order,
\begin{equation}
    \delta \Omega_\mu = \frac{1}{4} 
    \delta \omega_{\mu}^{\ ab} \gamma_{[a} \gamma_{b]},
\end{equation}
and the perturbed Dirac equation becomes
\begin{equation}
    i \gamma^\mu \left( \partial_\mu + \bar{\Omega}_\mu \right) \psi
    - m\psi
    = - i \gamma^\mu \delta \Omega_\mu \psi.
    \label{eq:dirac_noise}
\end{equation}

The right-hand side acts as an \emph{environmental interaction term}.  
Tracing over $h_{\mu\nu}$ in later sections produces a Lindblad operator governing decoherence.

%%%%%%%%%%%%%%%%%%%%%%%%%%%%%%%%%%%%%%%%%%%%%%%%%%%%%%%%%%%%
\section{Limitations and Open Issues}
%%%%%%%%%%%%%%%%%%%%%%%%%%%%%%%%%%%%%%%%%%%%%%%%%%%%%%%%%%%%

While the framework developed in this work provides a controlled
effective description of neutrino flavor evolution in curved spacetime,
several limitations and open theoretical issues should be emphasized.

First, the present analysis employs the standard Born--Markov and
secular approximations in deriving the Lindblad master equation.
Consequently, possible non-Markovian memory effects associated with
long-time gravitational correlations are neglected. In strongly
dynamical or highly fluctuating spacetime backgrounds, such effects
could generate departures from the local-in-time evolution considered
here.

Second, we neglect gravitational backreaction from the neutrino sector
onto the background geometry. The spacetime metric is therefore treated
as a fixed classical background, and the neutrino energy density is
assumed to remain sufficiently small that it does not modify the
Einstein equations. A fully self-consistent treatment incorporating
metric backreaction would require a significantly more complicated
semiclassical or quantum-gravitational analysis.

Third, the effective-field-theory operators introduced in
Sec.~3 should be interpreted as phenomenological curvature couplings
valid only below the ultraviolet cutoff scale $\Lambda$. The EFT
description is therefore expected to break down in regions where local
curvature invariants approach Planckian scales or where higher-order
operators become nonperturbatively important. In particular, we do not
attempt to extrapolate the formalism into regimes requiring a complete
theory of quantum gravity.

In addition, the Hawking-atmosphere model employed here should be viewed
primarily as an analytically tractable toy environment for generating
spin-connection fluctuations. The analysis does not assume that Hawking
radiation necessarily provides the dominant physical source of
astrophysical neutrino decoherence in realistic black-hole systems.

On the phenomenological side, the detector sensitivity estimates shown
in Sec.~7 are intended only as order-of-magnitude benchmarks rather
than full detector-level simulations. A more realistic observational
analysis would require detailed modeling of detector response,
background rejection, flavor tagging efficiencies, and astrophysical
source uncertainties.

Finally, the present work focuses primarily on stationary
Schwarzschild and Kerr geometries. Extensions to dynamical spacetimes,
binary mergers, cosmological backgrounds, and fully covariant
non-equilibrium quantum-gravity environments remain important open
directions for future investigation.

Despite these limitations, the framework developed here provides a
useful effective description connecting neutrino oscillations,
curved-spacetime dynamics, and open quantum systems in a form suitable
for quantitative phenomenological exploration.

%%%%%%%%%%%%%%%%%%%%%%%%%%%%%%%%%%%%%%%%%%%%%%%%%%%%%%%%%%%%
\section{Conclusions}
%%%%%%%%%%%%%%%%%%%%%%%%%%%%%%%%%%%%%%%%%%%%%%%%%%%%%%%%%%%%

In this work we developed an effective open-quantum-system framework
for neutrino flavor evolution in strong gravitational fields,
combining curved-spacetime neutrino propagation, spin--curvature
couplings, Kerr frame dragging, and gravitationally induced
decoherence within a unified formalism.

Starting from the Dirac equation in curved spacetime using the
vierbein and spin-connection formalism, we derived an effective
flavor Hamiltonian incorporating gravitational redshift effects,
rotation-induced spin couplings, and curvature-dependent corrections
to neutrino propagation. In contrast to many previous analyses that
focus only on gravitational phase shifts or purely phenomenological
decoherence parametrizations, the present framework connects the
dissipative sector directly to microscopic spin-connection
fluctuations through an explicit Lindblad construction.

A central result of this work is the derivation of curvature-induced
decoherence rates from spin-connection correlation functions in a
stationary stochastic gravitational environment. This provides a more
microscopic geometric interpretation of neutrino decoherence than
standard phenomenological damping models frequently adopted in the
literature. In particular, the formalism demonstrates how local
curvature invariants and frame-dragging effects can modify flavor
coherence during neutrino propagation near compact objects.

We further showed that Kerr rotation generates nontrivial distortions
in neutrino flavor evolution relative to the Schwarzschild case.
Rapidly rotating backgrounds produce enhanced flavor asymmetries,
energy-dependent oscillation distortions, and modified survival
probabilities in the strong-gravity regime. Low-energy neutrinos
experience the largest deviations, while high-energy neutrinos
asymptotically recover the standard flat-spacetime behavior.

Beyond oscillation probabilities, we investigated flavor-ratio
distortions and coherence loss in the language of quantum information.
The analysis of flavor-triangle evolution and entropy generation
demonstrates that gravitational decoherence can induce measurable
departures from the canonical democratic flavor composition expected
for astrophysical neutrinos. This establishes a direct conceptual
connection between neutrino oscillations, spacetime geometry, and
quantum-information observables.

An additional important aspect of the present work is the explicit
connection to detector-level phenomenology. We computed energy-dependent
event-rate distortions for neutrinos propagating in Schwarzschild and
Kerr backgrounds and compared the resulting signals with projected
sensitivities of IceCube-Gen2, KM3NeT, and P-ONE. The results indicate
that strong-gravity environments surrounding rapidly rotating compact
objects may generate potentially observable flavor distortions in
future high-energy neutrino experiments.

The framework developed here differs from much of the existing
literature in several important ways. In contrast to standard
phenomenological decoherence models, the dissipative sector in the
present analysis is derived directly from spin-connection fluctuations,
thereby establishing a more microscopic geometric origin for neutrino
decoherence in curved spacetime. Moreover, Kerr frame dragging,
spin--curvature couplings, and gravitationally induced decoherence are
treated simultaneously within a unified effective Hamiltonian
description rather than as isolated effects. The analysis also extends
beyond conventional oscillation probabilities by incorporating entropy
generation, coherence loss, and flavor-triangle evolution in the
language of quantum information. Finally, the formalism is connected
explicitly to detector-level phenomenology through energy-dependent
event-rate distortions and comparisons with projected sensitivities of
next-generation neutrino telescopes such as IceCube-Gen2, KM3NeT, and
P-ONE.
```

Although the present analysis remains within an effective-field-theory
and weak-coupling regime, the results suggest that neutrino flavor
evolution may provide a potentially sensitive probe of spacetime
structure in strong gravitational environments. More broadly, the
formalism developed here illustrates how neutrino oscillations can
serve as a bridge connecting curved-spacetime quantum field theory,
open quantum systems, astrophysical phenomenology, and quantum
information dynamics.

Future extensions of this framework may include fully non-Markovian
gravitational environments, dynamical spacetimes associated with
compact-object mergers, neutrino propagation in cosmological
backgrounds, and possible connections with more fundamental theories
of quantum gravity.

We therefore conclude that neutrino flavor evolution near compact
gravitational sources provides a theoretically rich and potentially
observable arena for studying the interplay between quantum coherence,
curved spacetime, and strong-gravity astrophysics.

%%%%%%%%%%%%%%%%%%%%%%%%%%%%%%%%%%%%%%%%%%%%%%%%%%%%%%%%%%
\appendix
\label{app:cp}
\section{Complete Positivity of the Curvature-Induced
Lindblad Generator}
%%%%%%%%%%%%%%%%%%%%%%%%%%%%%%%%%%%%%%%%%%%%%%%%%%%%%%%%%%

In this appendix we derive the explicit Lindblad operators
associated with curvature-induced spin-connection
fluctuations and demonstrate the complete positivity of the
resulting quantum dynamical semigroup.

The interaction Hamiltonian between the neutrino and the
stochastic gravitational environment is

\begin{equation}
H_{\rm int}(t)
=
\sum_i
A_i \otimes B_i(t),
\label{eq:A1}
\end{equation}

where the system operators are identified with the spin
generators

\begin{equation}
A_i
=
\Sigma_i ,
\end{equation}

while the environmental operators are constructed from the
fluctuating spin connection,

\begin{equation}
B_i(t)
=
\delta\Omega_i(t).
\end{equation}

Within the Born--Markov approximation the reduced density
matrix obeys

\begin{equation}
\frac{d\rho}{dt}
=
-i[H_{\rm eff},\rho]
+
\mathcal{D}[\rho],
\end{equation}

with dissipator

\begin{equation}
\mathcal{D}[\rho]
=
\sum_{ij}
C_{ij}
\left(
A_j \rho A_i^\dagger
-
\frac12
\{A_i^\dagger A_j,\rho\}
\right).
\label{eq:A2}
\end{equation}

The Kossakowski matrix is obtained from the
spin-connection correlation functions,

\begin{equation}
C_{ij}
=
\int_0^\infty ds\,
\Big\langle
B_i(s)B_j(0)
\Big\rangle .
\label{eq:A3}
\end{equation}

%%%%%%%%%%%%%%%%%%%%%%%%%%%%%%%%%%%%%%%%%%%%%%%%%%%%%%%%%%%%
%%%%%%%%%%%%%%%%%%%%%%% APPENDICES %%%%%%%%%%%%%%%%%%%%%%%%%
%%%%%%%%%%%%%%%%%%%%%%%%%%%%%%%%%%%%%%%%%%%%%%%%%%%%%%%%%%%%

\appendix

%%%%%%%%%%%%%%%%%%%%%%%%%%%%%%%%%%%%%%%%%%%%%%%%%%%%%%%%%%%%
\section{Spin Connection and Kerr Tetrad Derivation}
%%%%%%%%%%%%%%%%%%%%%%%%%%%%%%%%%%%%%%%%%%%%%%%%%%%%%%%%%%%%

In this appendix we present the derivation of the spin connection and
curved-space Dirac structure employed in the main text for neutrino
propagation in Kerr spacetime.

The Kerr metric in Boyer--Lindquist coordinates is
\begin{equation}
ds^2
=
-\left(1-\frac{2GMr}{\Sigma}\right)dt^2
-\frac{4GMar\sin^2\theta}{\Sigma}dt\,d\phi
+\frac{\Sigma}{\Delta}dr^2
+\Sigma d\theta^2
+\frac{A\sin^2\theta}{\Sigma}d\phi^2,
\end{equation}
where
\begin{equation}
\Delta = r^2-2GMr+a^2,
\qquad
\Sigma = r^2+a^2\cos^2\theta,
\end{equation}
and
\begin{equation}
A=(r^2+a^2)^2-a^2\Delta\sin^2\theta.
\end{equation}

A convenient orthonormal tetrad basis is
\begin{align}
e^{(0)}
&=
\sqrt{\frac{\Delta}{\Sigma}}
(dt-a\sin^2\theta\,d\phi),
\\
e^{(1)}
&=
\sqrt{\frac{\Sigma}{\Delta}}\,dr,
\\
e^{(2)}
&=
\sqrt{\Sigma}\,d\theta,
\\
e^{(3)}
&=
\frac{\sin\theta}{\sqrt{\Sigma}}
[(r^2+a^2)d\phi-a\,dt].
\end{align}

The spin connection is obtained from
\begin{equation}
\omega_{\mu}^{ab}
=
e^a_{\nu}
\left(
\partial_\mu e^{b\nu}
+
\Gamma^\nu_{\mu\lambda}e^{b\lambda}
\right).
\end{equation}

The curved-space gamma matrices satisfy
\begin{equation}
\gamma^\mu(x)
=
e^\mu_a(x)\gamma^a,
\end{equation}
with
\begin{equation}
\{\gamma^a,\gamma^b\}
=
2\eta^{ab}.
\end{equation}

The spinor covariant derivative becomes
\begin{equation}
D_\mu
=
\partial_\mu
+
\Omega_\mu,
\end{equation}
where
\begin{equation}
\Omega_\mu
=
\frac14
\omega_\mu^{ab}\gamma_{[a}\gamma_{b]}.
\end{equation}

Expanding in the slow-rotation regime $(a/r\ll1)$ yields the leading
frame-dragging contribution
\begin{equation}
\Omega_\phi
\simeq
\frac{aGM}{r^3}\Sigma^{03},
\end{equation}
which generates the effective Kerr spin-coupling Hamiltonian used in
Sec.~3.

%%%%%%%%%%%%%%%%%%%%%%%%%%%%%%%%%%%%%%%%%%%%%%%%%%%%%%%%%%%%
\section{Born--Markov--Secular Derivation of the Lindblad Equation}
%%%%%%%%%%%%%%%%%%%%%%%%%%%%%%%%%%%%%%%%%%%%%%%%%%%%%%%%%%%%

In this appendix we derive the reduced open-system evolution equation
for neutrino flavor propagation in the presence of stochastic
spin-connection fluctuations.

The total Hamiltonian is written as
\begin{equation}
H_{\rm tot}
=
H_\nu
+
H_{\rm env}
+
H_{\rm int},
\end{equation}
with interaction Hamiltonian
\begin{equation}
H_{\rm int}(t)
=
\sum_\alpha
A_\alpha(t)\otimes B_\alpha(t),
\end{equation}
where $A_\alpha$ act on the neutrino Hilbert space and
$B_\alpha$ represent gravitational environmental operators.

In the interaction picture,
\begin{equation}
\dot{\rho}_{\rm tot}(t)
=
-i[H_{\rm int}(t),\rho_{\rm tot}(t)].
\end{equation}

Iterating once and tracing over environmental degrees of freedom gives
\begin{align}
\dot{\rho}_\nu(t)
=
-
\int_0^\infty ds
\,{\rm Tr}_{\rm env}
\Big[
H_{\rm int}(t),
[H_{\rm int}(t-s),
\rho_\nu(t)\otimes\rho_{\rm env}]
\Big].
\end{align}

Assuming:
\begin{enumerate}
\item weak coupling (Born approximation),
\item short environmental correlation time (Markov limit),
\item rapidly oscillating off-diagonal terms (secular approximation),
\end{enumerate}
the reduced density matrix obeys the Lindblad equation
\begin{equation}
\dot{\rho}_\nu
=
-i[H_{\rm eff},\rho_\nu]
+
\sum_i
\left(
L_i\rho_\nu L_i^\dagger
-
\frac12
\{L_i^\dagger L_i,\rho_\nu\}
\right).
\end{equation}

For exponential spin-connection correlations,
\begin{equation}
G(s)
=
\alpha e^{-|s|/\tau_c},
\end{equation}
the dissipative coefficient becomes
\begin{equation}
\Gamma_{\rm grav}
=
\int_0^\infty ds\,G(s)
=
\alpha\tau_c.
\end{equation}

%%%%%%%%%%%%%%%%%%%%%%%%%%%%%%%%%%%%%%%%%%%%%%%%%%%%%%%%%%%%
\section{Positivity of the Kossakowski Matrix}
%%%%%%%%%%%%%%%%%%%%%%%%%%%%%%%%%%%%%%%%%%%%%%%%%%%%%%%%%%%%

Complete positivity of the reduced density-matrix evolution requires
the Kossakowski matrix to be positive semidefinite.

The dissipative sector may be written as
\begin{equation}
\mathcal{D}[\rho]
=
\sum_{ij}
C_{ij}
\left(
F_i\rho F_j^\dagger
-
\frac12
\{F_j^\dagger F_i,\rho\}
\right),
\end{equation}
where $C_{ij}$ is the Kossakowski matrix.

For stationary Gaussian spin-connection fluctuations,
\begin{equation}
C_{ij}
=
\int_0^\infty ds
\,
\langle
B_i(s)B_j(0)
\rangle.
\end{equation}

The correlator matrix is Hermitian,
\begin{equation}
C_{ij}=C_{ji}^*,
\end{equation}
and positivity follows from
\begin{equation}
\sum_{ij}
v_i^* C_{ij} v_j
=
\int_0^\infty ds
\,
\left\langle
X^\dagger(s)X(0)
\right\rangle
\ge0,
\end{equation}
where
\begin{equation}
X=\sum_i v_i B_i.
\end{equation}

Hence all eigenvalues $\lambda_i$ satisfy
\begin{equation}
\lambda_i\ge0,
\end{equation}
guaranteeing completely positive trace-preserving evolution.

%%%%%%%%%%%%%%%%%%%%%%%%%%%%%%%%%%%%%%%%%%%%%%%%%%%%%%%%%%%%
\section{Benchmark Parameters and Numerical Inputs}
%%%%%%%%%%%%%%%%%%%%%%%%%%%%%%%%%%%%%%%%%%%%%%%%%%%%%%%%%%%%

The numerical analysis presented in Secs.~7--8 employs the following
benchmark oscillation and gravitational parameters:
\begin{align}
\Delta m^2
&=
2.5\times10^{-3}\ {\rm eV}^2,
\\
\theta
&=
\frac{\pi}{4},
\\
a/M
&=
\{0,0.3,0.6,0.95\}.
\end{align}

The neutrino energies are chosen in the range
\begin{equation}
1~{\rm GeV}
\le E \le
10^8~{\rm GeV},
\end{equation}
covering the sensitivity windows of current and future neutrino
telescopes.

The decoherence parameter is parametrized as
\begin{equation}
\Gamma_{\rm grav}(r)
=
\gamma_0
\frac{GM}{r^3},
\end{equation}
with $\gamma_0$ treated phenomenologically.

Throughout the numerical analysis we employ natural units
\begin{equation}
c=\hbar=G=1.
\end{equation}

%%%%%%%%%%%%%%%%%%%%%%%%%%%%%%%%%%%%%%%%%%%%%%%%%%%%%%%%%%%%
\section{Detector Sensitivity Modeling}
%%%%%%%%%%%%%%%%%%%%%%%%%%%%%%%%%%%%%%%%%%%%%%%%%%%%%%%%%%%%

To estimate the potential observability of curvature-induced neutrino
distortions, we compare the predicted event-rate modification
\begin{equation}
\delta N(E)
=
\frac{
N_{\rm Kerr}(E)-N_{\rm Schw}(E)
}{
N_{\rm Schw}(E)
}
\end{equation}
with projected sensitivities of next-generation neutrino telescopes.

The event rate is modeled as
\begin{equation}
N(E)
=
\Phi(E)\,
\sigma_{\nu N}(E)\,
A_{\rm eff}(E)\,
T,
\end{equation}
where $\Phi(E)$ denotes the neutrino flux,
$\sigma_{\nu N}$ is the neutrino--nucleon cross section,
$A_{\rm eff}(E)$ is the detector effective area,
and $T$ is the exposure time.

Projected sensitivity regions are estimated using published detector
performance studies for IceCube-Gen2, KM3NeT, and P-ONE.
The shaded bands and horizontal sensitivity thresholds shown in
Fig.~4 correspond to approximate projected $1\sigma$ statistical
reach estimates for these experiments.

The detector overlays are intended as phenomenological benchmarks
illustrating the potential observability of curvature-induced effects
rather than as full detector-level simulations.
```

%%%%%%%%%%%%%%%%%%%%%%%%%%%%%%%%%%%%%%%%%%%%%%%%%%%%%%%%%%%%%%%%%%%%%%%%

%%%%%%%%%%%%%%%%%%%%%%%%%%%%%%%%%%%%%%%%%%%%%%%%%%%%%%%%%%%%%%%%%%%%%%%%

\end{document}